\documentclass[final,11pt]{elsarticle}

\usepackage[english]{babel}
\usepackage{textcomp,color}
\usepackage{graphicx}
\usepackage{amsmath}
\usepackage{amssymb}
\usepackage{amsxtra}
\usepackage{mathrsfs}
\usepackage[left=2cm,right=2cm,top=2cm,bottom=2cm,bindingoffset=0cm]{geometry}

\usepackage {float}

\usepackage{natbib}

\usepackage{tikz}
\usepackage[utf8]{inputenc}
\usepackage{feynmf}
\usepackage{cancel}
\usepackage{indentfirst}

\renewcommand{\Im}{\mathop{\mathrm{Im}}\nolimits}

\newcommand{\cosech}{\mathop{\mathrm{cosech}}\nolimits}

\usepackage[normalem]{ulem}
\usepackage{simpler-wick}

\usepackage{graphicx,xcolor}

\makeatletter
\def\Appendix{\appendix
\def\@seccntformat##1{Appendix~\csname the##1\endcsname.~~}}
\makeatother
\makeatletter
\@addtoreset{equation}{section}

\makeatother

\def\XXint#1#2#3{{\setbox0=\hbox{$#1{#2#3}{\int}$}
\vcenter{\hbox{$#2#3$}}\kern-.5\wd0}}

\journal{Nuclear Physics B}

\begin{document}

\begin{frontmatter}



\title{On bosonic Thirring model in Minkowski signature}

\author[first,second]{Mikhail Alfimov\corref{cor1}}
\ead{malfimov@hse.ru}
\cortext[cor1]{Corresponding author. Telephone number: +7 (903) 819-33-08.}
\author[first]{Andrey Kurakin}
\ead{aakurakin@hse.ru}
\affiliation[first]{organization={HSE University},
            addressline={ul. Usacheva, d. 6}, 
            city={Moscow},
            postcode={119048},
            country={Russia}}

\affiliation[second]{organization={P.N. Lebedev Physical Institute of the Russian Academy of Sciences},
            addressline={Leninskiy pr., d. 53},
            city={Moscow},
            postcode={119991},
            country={Russia}}

\begin{abstract}
We present the way to continue the bosonic Thirring model or $\beta\gamma$-system with quartic interaction to Minkowski signature, based on the symmetries of this model. It is shown that the considered Minkowski version of the model is one-loop renormalizable. Based on this, we find the amplitudes of scattering of the excitations corresponding to the $\gamma$ and $\bar{\gamma}$ fields up to the one-loop order. In particular, it was computed that the $2 \rightarrow 2$ amplitudes of these excitations possess property of reflectionless scattering and thus the corresponding $S$-matrix of such excitations satisfies the Yang-Baxter equation. The obtained $S$-matrix elements for $\gamma$ and $\bar{\gamma}$ are shown to coincide with the corresponding $S$-matrix elements of the solitons in the complex sine-Gordon model proposed by Dorey and Hollowood.
\end{abstract}




\end{frontmatter}

\section{Introduction}

Dual description of deformed $OSp(N|2m)$ sigma models in terms of Toda-like theories, which are determined by the system of the so-called screening charges \cite{Wakimoto:1986gf}, was studied in the recent work \cite{Alfimov:2020jpy} by the one of the authors. The key feature of such description is that the $S$-matrix calculated from the dual Toda-like Lagrangian order by order in perturbation theory coincides with the corresponding sigma model $S$-matrix, which represents a deformed solution of the Yang-Baxter equation. This description was constructed for several cases, such as $O(N)$ \cite{Fateev:2018yos, Litvinov:2018bou} or $\mathbb{C}P^{N}$ \cite{Litvinov:2019rlv} deformed sigma models\footnote{For the $O(3)$ case it was first done by Alexey Zamolodchikov (unpublished).}. However, compared to these cases, there are some novel pecularities in the $OSp(N|2m)$ case. Namely, in the dual Lagrangians of the $\eta$-deformed $OSp(5|2)$ and $OSp(7|2)$ sigma models because of the presence of odd coordinates in the target space there appears the so-called bosonic Thirring model, whose Euclidean action is determined by the Lagrangian of $\beta\gamma$-system with the quartic interaction
\begin{equation}\label{Euclidean_betagamma_Lagrangian}
    \mathcal{L}_{\beta\gamma}^{(E)}= \bar{\beta}\partial\bar{\gamma}+\beta\bar{\partial}\gamma-m\bar{\beta}\beta-m\bar{\gamma}\gamma+g^2\beta\bar{\beta}\gamma\bar{\gamma}\,.
\end{equation}
As it was said above, one of the ways of confirming the validity of dual description is to compare the $S$-matrix calculated perturbatively from the dual Toda-like theory with the exact $S$-matrix originating from the symmetries of the corresponding sigma model (for the examples see \cite{Litvinov:2018bou}). However, in contrast to the fermionic version of the Thirring model, the analogue of \eqref{Euclidean_betagamma_Lagrangian} in Minkowski signature, which allows to correctly reproduce the $S$-matrix for the corresponding $OSp(N|2m)$ sigma model, is not known. The objective of the present article is to find such an analogue in the bosonic case.

Let us first present the information about the tree level. As it was pointed out in \cite{Alfimov:2020jpy}, in order to calculate the $S$-matrix for the excitations corresponding to the fields $\gamma$ and $\bar{\gamma}$, we can take the Gaussian integral over the fields $\beta$ and $\bar{\beta}$ (see the details in the~\ref{app_Path_integral}). After this procedure we obtain the classical Lagrangian of complex sine-Gordon (CSG) theory \cite{Pohlmeyer:1975nb,Lund:1976ze,Lund:1977dt,Getmanov:1977hk} up to the contribution of some functional determinant
\begin{equation}\label{Euclidean_CsG_Lagrangian}
    \mathcal{L}_{CsG}^{(E)}=\frac{|\partial\psi|^2}{1-g^2|\psi|^2}-m^2|\psi|^2\,,
\end{equation}
which can be straightforwardly continued to Minkowski signature
\begin{equation}\label{Minkowskian_CsG_Lagrangian}
    \mathcal{L}_{CsG}^{(M)}=\frac{1}{2}\left(\frac{\partial_+ \psi_a \partial_- \psi_a}{1-\frac{g^2 \psi_a^2}{2}}-m^2 \psi_a^2\right)\,, \quad a=1,2\,,
\end{equation}
where $\psi=(\psi_1+i\psi_2)/\sqrt{2}$. In the article \cite{deVega:1981ka} there was conducted perturbative analysis of the theory \eqref{Minkowskian_CsG_Lagrangian}. One-loop $S$-matrix for this Lagrangian \eqref{Minkowskian_CsG_Lagrangian} was calculated, however, the theory was shown to be not one-loop renormalizable. Upon supplementing the theory with the proper counterterm, it becomes one-loop renormalizable and its one-loop $S$-matrix starts to satisfy the Yang-Baxter equation, which is equivalent to the condition of reflectionless scattering in this case. Then, in the work \cite{Dorey:1994mg} the semiclassical spectrum and the $S$-matrix of the theory \eqref{Minkowskian_CsG_Lagrangian} was studied. A hypothesis was put forward for the minimal exact $S$-matrix \cite{Zamolodchikov:1978xm} for specific values of coupling constant of such a theory consistent with the semiclassical spectrum of the theory.

Our hypothesis is that the theory \eqref{Euclidean_betagamma_Lagrangian}, properly continued to the Minkowski signature, yields the same $S$-matrix for the excitations corresponding to the fields $\gamma$ and $\bar{\gamma}$ as proposed in \cite{Dorey:1994mg} order by order in the coupling constant $g$. The first step to achieve this is to understand how to interpret the Lagrangian \eqref{Euclidean_betagamma_Lagrangian} in Minkowski spacetime. We offer a way to do this based on the spacetime and global symmetries of the action \eqref{Euclidean_betagamma_Lagrangian}. The main idea is to look for the $SO(2)$ subgroup of the full symmetry group of the theory and turn it into the $SO(1,1)$, thus passing to the Lorentz group symmetry. After doing this the next important question, as we are interested in loop calculations, is the analysis of divergences. To understand the renormalization of the  fields, mass and coupling constant, we calculate two- and four-point functions, establishing which counterterms are needed at one-loop order and partially at two-loop order.

The next step is to calculate the $S$-matrix of the continuation of the theory \eqref{Euclidean_betagamma_Lagrangian} to the Minkowski signature. This is done by applying the Lehmann-Symanczik-Zimmermann (LSZ) reduction formula \cite{Lehmann:1954rq} to the four-point correlation function of the fields $\gamma$ and $\bar{\gamma}$, which yields the desired amplitudes. We were able to show the coincidence our result with the amplitudes from \cite{Dorey:1994mg} at the tree and $1$-loop level, thus confirming our hypothesis at these orders.

The article is organized as follows. In the Section~\ref{CsG_setting} we remind the reader facts that are known about the complex sine-Gordon theory regarding its spectrum and scattering. In the next Section~\ref{betagamma_system_continuation} we explain the way we use to continue the theory \eqref{Euclidean_betagamma_Lagrangian} to the Minkowski signature and setting the propagators and vertices of such a theory and also show how to obtain the $\beta\gamma$-system in Minkowski signature from the gauged Wess-Zumino-Witten model with integrable potential. In the Section~\ref{Two_point_functions} we calculate the two-point functions at one and two loops and also do some analysis of necessary counterterms. Then, in the Section~\ref{Four_point_functions} we turn to the four-point function of the fields $\gamma$ and $\bar{\gamma}$, from which we extract the $2 \rightarrow 2$ amplitude to compare with the $S$-matrix of the CSG theory \cite{Dorey:1994mg}. In the Section~\ref{Conclusions} we make concluding remarks and discuss further open questions.

\section{Complex sine-Gordon theory}\label{CsG_setting}

Let us start from refreshing our knowledge about the spectrum and scattering in the CSG theory \cite{Pohlmeyer:1975nb,Lund:1976ze,Lund:1977dt,Getmanov:1977hk}. In this section we are going to remember the mass spectrum of the CSG theory and write down the all-loop $S$-matrix, which was conjectured earlier in \cite{Dorey:1994mg}. First we write the weak coupling expansion of the CSG classical Lagrangian \eqref{Minkowskian_CsG_Lagrangian}, which is\footnote{Notice that in \cite{Hoare:2010fb} the authors use different coupling $\mathbf{g}^2$, which is related to ours by $g^2=2\mathbf{g}^2$.}
\begin{equation}\label{eq:lagrSG}
    \mathcal{L}_{CsG} = \frac{1}{2}\partial_+\psi_a \partial_-\psi_a -\frac{m^2}{2}\psi_a\psi_a + \frac{1}{2}\sum\limits_{k=1}^{+\infty}\left(\frac{g^2}{2}\right)^k\partial_+\psi_a\partial_-\psi_a (\psi_b\psi_b)^k\,.
\end{equation}
In this theory we have the solitons as elementary excitations, which are classified by the charge \cite{deVega:1982sh}. Soliton charge is defined modulo $2\pi/g^2$
\begin{equation}\label{soliton_charges}
    Q=\pm 1,\ldots, Q_{max}=\pm\left[\frac{2\pi}{g^2}\right]\,.
\end{equation}
It was shown in \cite{deVega:1982sh}, that these solitons have the following mass at the semiclassical level
\begin{equation}\label{mass_spectrum}
    M(Q) = \frac{4m}{g^2}\left|\sin\left(\frac{g^2Q}{4}\right)\right|\,.
\end{equation}
Also in the same article it was conjectured, that under finite renormalization of the coupling, which corresponds to adding proper counterterms to the action \eqref{eq:lagrSG}
\begin{align}\label{coupling_constant_renormalization}
    g_R^2=\frac{g^2}{1-\frac{g^2}{4\pi}}
\end{align}
analogous to the sine-Gordon (SG) theory, the semiclassical mass spectrum \eqref{mass_spectrum} becomes exact in the coupling constant. Given this information, we can now turn to the $S$-matrix of such soliton excitations.

In \cite{deVega:1981ka} there was computed the one-loop $S$-matrix, which for elementary excitations of the fields $\psi_a$ is given by the following expression
\begin{equation}\label{CsG_S-matrix}
    S_{ij}^{i'j'}(\theta) = S_1(\theta)\delta_{ij}\delta^{i'j'} +S_2(\theta)\delta_{i}^{i'}\delta_{j}^{j'}+S_3(\theta)\delta_{i}^{j'}\delta_{j}^{i'}\,,
\end{equation}
where all the indices take the values $1$ and $2$ and the functions $S_{1,2,3}(\theta)$ are
\begin{align}\label{S123_bare}
    & S_1(\theta) = -\frac{ig^2}{2}\coth\theta+i\frac{g^4}{8\pi}(\cosech\theta+\coth\theta)+\frac{g^4}{4}\coth\theta\cosech\theta +\mathcal{O}(g^6)\,, \\
    & S_2(\theta) = 1+\frac{ig^2}{2}\cosech\theta-\frac{g^4}{4}\left(\frac{1}{2}+\cosech^2\theta\right)+\mathcal{O}(g^6)\,, \notag \\
    & S_3(\theta) =\frac{ig^2}{2}\coth\theta+i\frac{g^4}{8\pi}(\cosech\theta-\coth\theta)-\frac{g^4}{4}\coth\theta\cosech\theta+\mathcal{O}(g^6)\,. \notag
\end{align}
However, the $S$-matrix determined by \eqref{CsG_S-matrix} and \eqref{S123_bare} does not satisfy the Yang-Baxter equation, which in the case in question is equivalent to the condition of reflectionless scattering
\begin{equation}
S_1(\theta)+S_3(\theta)=0\,.
\end{equation}
As it was shown in \cite{deVega:1981ka} if we supplement the action with the counterterm
\begin{equation}\label{DVM_countertetm}
\Delta\mathcal{L}=\frac{m^2 g^4}{16\pi}\psi_a \psi_a \psi_b \psi_b\,,    
\end{equation}
the one-loop $S$-matrix takes the form
\begin{align}\label{S123}
    & S_1(\theta) = -\frac{ig^2}{2}\coth\theta+i\frac{g^4}{8\pi}\coth\theta+\frac{g^4}{4}\coth\theta\cosech\theta +\mathcal{O}(g^6)\,, \\
    & S_2(\theta) = 1+\frac{ig^2}{2}\cosech\theta-i\frac{g^4}{8\pi}\cosech\theta-\frac{g^4}{4}\left(\frac{1}{2}+\cosech^2\theta\right)+\mathcal{O}(g^6)\,, \notag \\
    & S_3(\theta) =\frac{ig^2}{2}\coth\theta-i\frac{g^4}{8\pi}\coth\theta-\frac{g^4}{4}\coth\theta\cosech\theta+\mathcal{O}(g^6)\,. \notag
\end{align}
It is known that the $S$-matrix \eqref{CsG_S-matrix} with \eqref{S123} satisfies the Yang-Baxter equation up to one-loop order thanks to the reflectionless condition
\begin{equation}\label{reflectionless_condition}
    S_1(\theta)+S_3(\theta)=\mathcal{O}(g^6)\,.
\end{equation}

Let us change the basis to the complex one
\begin{equation}\label{gamma_complex_basis}
    \gamma=\frac{1}{\sqrt{2}}(\psi_1+i\psi_2)\,, \quad \bar{\gamma}=\frac{1}{\sqrt{2}}(\psi_1-i\psi_2)\,.
\end{equation}
In the complex basis \eqref{gamma_complex_basis} the formulas \eqref{S123} correspond exactly to the $2 \rightarrow 2$ scattering amplitudes of the particles $\gamma$ and $\bar{\gamma}$, which can be written as
\begin{align}\label{CsG_complex_basis}
        & S_{\gamma\gamma}^{\gamma\gamma}(\theta)=S_{\bar{\gamma}\bar{\gamma}}^{\bar{\gamma}\bar{\gamma}}(\theta)=S_2(\theta)+S_3(\theta)\,, \\
        & S_{\gamma\bar{\gamma}}^{\gamma\bar{\gamma}}(\theta)=S_1(\theta)+S_2(\theta)\,, \notag \\
        & S_{\gamma\bar{\gamma}}^{\bar{\gamma}\gamma}(\theta)=S_1(\theta)+S_3(\theta)\,. \notag
\end{align}
Given that at one loop the amplitude is reflectionless \eqref{reflectionless_condition}, by using the crossing symmetry we can express the whole one-loop $S$-matrix in terms of solely one function
\begin{equation}\label{eq:S-matrix}
    S(\theta) = S_1(\theta) + S_2(\theta) = 1-\frac{ig^2}{2}\tanh\frac{\theta}{2}+\frac{ig^4}{8\pi}\tanh\frac{\theta}{2}-\frac{g^4}{8}\tanh^2 \frac{\theta}{2}+\mathcal{O}(g^6)\,. 
\end{equation}

However, introduction of the counterterm \eqref{DVM_countertetm} is not the unique way to restore the symmetries responsible for the integrability of the model at the quantum level. As it was explained in \cite{Hoare:2010fb} by using the gauged Wess-Zumino-Witten (gWZW) model formulation of the CSG model \eqref{Minkowskian_CsG_Lagrangian} it is possible to produce the counterterns different from \eqref{DVM_countertetm}, which nevertheless lead to the $S$-matrix, which satisfies the Yang-Baxter equation. The authors of that work showed that integration of the gWZW action over the fields $A_{\pm}$ produces the functional determinant contribution, which, under the proper identification of the fields, yields the $S$-matrix we write down below up to one-loop order. The advantage of the gWZW or $\beta\gamma$-system approach we consider below is that the symmetries responsible or integrability are more manifest in this representation of the CSG model and we do not need to include additional counterterms if we do not integrate over gauge or $\beta$ fields respectively.

In relation to the spectrum given by \eqref{soliton_charges}, \eqref{mass_spectrum} and \eqref{coupling_constant_renormalization} the particles $\gamma$ and $\bar{\gamma}$ correspond to the solitons with the charges $Q=1$ and $Q=-1$ respectively. In \cite{Dorey:1994mg} there was proposed exact $2 \rightarrow 2$ $S$-matrix of a pair of solitons with the charges $Q_1$ and $Q_2$ denoted as $S_{Q_1,Q_2}(\theta)$. We are interested in the case $Q_1=1$, $Q_2=-1$ exact S-matrix
\begin{equation}\label{eq:Seaxt1}
    S_{1,-1}(\theta)=S_{1,1}(i\pi-\theta)= \frac{\sinh\left(\frac{i\pi}{2}-\frac{\theta}{2}+\frac{ig_R^2}{4}\right)}{\sinh\left(\frac{i\pi}{2}-\frac{\theta}{2}-\frac{ig_R^2}{4}\right)}\,.
\end{equation}
where the $g_R$ is determined by \eqref{coupling_constant_renormalization} and is subject to the condition
\begin{equation}\label{coupling_quantization}
g_R^2=\frac{4\pi}{\mathrm{k}}\, \quad \mathrm{k} \in \mathbb{N}\,.
\end{equation}
Expanding \eqref{eq:Seaxt1} into the series up to the terms of the order $g^4$ we obtain
\begin{equation}\label{eq:Seaxt2}
    S_{1,-1}(\theta)=1-\frac{ig^2}{2}\tanh\frac{\theta}{2}-\frac{ig^4}{8\pi}\tanh\frac{\theta}{2}-\frac{g^4}{8}\tanh^2 \frac{\theta}{2}+\mathcal{O}(g^6)\,,
\end{equation}
which in our terms corresponds to the amplitude $S_{\gamma\bar{\gamma}}^{\gamma\bar{\gamma}}(\theta)$. One can notice the discrepancy at the order $g^4$ in \eqref{eq:S-matrix} and \eqref{eq:Seaxt2}, which, as commented in \cite{Hoare:2010fb}, can be explained by the choice of the renormalization scheme. We are to compare the results \eqref{eq:S-matrix} and \eqref{eq:Seaxt2} to the $2 \rightarrow 2$ amplitude, obtained from the version of the theory \eqref{Euclidean_betagamma_Lagrangian} properly analytically continued to Minkowski signature. We are going to conduct this in the next Section.

\section{$\beta\gamma$-system in Minkowski space}\label{betagamma_system_continuation}

Our objective now is to find the way to define the bosonic Thirring model \eqref{Euclidean_betagamma_Lagrangian} in Minkowski signature. Contrary to bosonic case, fermionic Thirring model in Minkowski signature is known \cite{Korepin:1979qq}. The main idea is to analyze the symmetries of the model \eqref{Euclidean_betagamma_Lagrangian} and, based on that find the $SO(2)$ subgroup, which can be continued to the Lorentz group $SO(1,1)$. Next subsection will tell us about the symmetries of \eqref{Euclidean_betagamma_Lagrangian}.

\subsection{Symmetries and continuation to Minkowski signature}

Analogous to the Coleman-Mandelstam duality \cite{Coleman:1974bu} in \cite{Alfimov:2020jpy} it was shown that at least at the classical level the CSG model is equivalent to the bosonic version of the Thirring model, which is determined by the Lagrangian of the $\beta\gamma$-system with quartic interaction
\begin{equation}
    \mathcal{L}_{T} = i\mathcal{\bar B}\gamma^{\mu}\partial_{\mu}\mathcal{B} - m\mathcal{\bar B}\mathcal{B} - \frac{\pi\lambda^2}{2}(\mathcal{\bar B}\gamma^{\mu}\mathcal{B})^2,
\end{equation}
where
\begin{equation}
\gamma^1 = \sigma_1\,, \quad \gamma^2 = \sigma_2\,, \quad \mathcal{\bar B} = \left( \begin{matrix} -i\bar\beta & \gamma	\end{matrix}\right)\,, \quad \mathcal{B} = \left( \begin{matrix} i\beta \\ \bar{\gamma} \end{matrix}\right)\,.
\end{equation}
 Then in terms of $\beta$ and $\gamma$ fields we get the Lagrangian
\begin{equation}\label{Euclidean_beta_gamma_action}
\mathcal{L}_{\beta\gamma}=\beta\bar{\partial}\gamma+\bar{\beta}\partial\bar{\gamma}-m \beta\bar{\beta}-m \gamma\bar{\gamma}-2\pi\lambda^2 \beta\bar{\beta}\gamma\bar{\gamma}\,.
\end{equation}
If we replace the coupling constant
\begin{equation}
    -2\pi\lambda^2=g^2\,,
\end{equation}
the $\beta\gamma$ Lagrangian in Euclidean space takes the same form as in \eqref{Euclidean_betagamma_Lagrangian}
\begin{equation}
\mathcal{L}_{\beta\gamma}=\beta\bar{\partial}\gamma+\bar{\beta}\partial\bar{\gamma}-m \beta\bar{\beta}-m \gamma\bar{\gamma}+g^2 \beta\bar{\beta}\gamma\bar{\gamma}\,.
\end{equation}

To understand the way of analytic continuation we have to first analyze the symmetries of the theory \eqref{Euclidean_betagamma_Lagrangian}. This theory has the $SO(2)$ symmetry corresponding to the global rotation of the fields $\beta$ and $\gamma$
\begin{equation}
    \beta \rightarrow e^{i\alpha}\beta\,, \quad \bar{\beta} \rightarrow e^{-i\alpha}\bar{\beta}\,, \quad \gamma \rightarrow e^{-i\alpha}\gamma\,, \quad \bar{\gamma} \rightarrow e^{i\alpha}\bar{\gamma}
\end{equation}
and the symmetry corresponding to the space $SO(2)$ transformation
\begin{align}
    & \partial \rightarrow e^{i\phi}\partial\,, \quad \beta \rightarrow e^{\frac{i\phi}{2}}\beta\,, \quad \gamma \rightarrow e^{\frac{i\phi}{2}}\gamma\,, \\
    & \bar{\partial} \rightarrow e^{-i\phi}\bar{\partial}\,, \quad \bar{\beta} \rightarrow e^{-\frac{i\phi}{2}}\bar\beta\,, \quad \bar{\gamma} \rightarrow e^{-\frac{i\phi}{2}}\bar{\gamma}\,. \notag
\end{align}
Thus, the total symmetry group of the $\beta\gamma$-system in Euclidean space is $SO(2) \times SO(2)$. We would like to define now this theory in Minkowski signature and the way to achieve this is to find the $SO(2)$ subgroup inside this $SO(2) \times SO(2)$ and then to turn it into the $2d$ Lorentz group $SO(1,1)$.

Let us select subgroup $SO(2)$ in the symmetry group $SO(2)\times SO(2)$  as follows
\begin{align}
    & \partial \rightarrow e^{i\phi}\partial\,, \quad \beta \rightarrow e^{i\phi}\beta\,, \quad \gamma \rightarrow \gamma\,, \\
    & \bar{\partial} \rightarrow e^{-i\phi}\bar{\partial}\,, \quad \bar{\beta} \rightarrow e^{-i\phi}\bar{\beta}\,, \quad \bar{\gamma} \rightarrow \bar{\gamma}\,. \notag
\end{align}
(we set $\alpha = \phi/2$ when choosing the subgroup). In this case the fields $\gamma$ and $\bar{\gamma}$ remain invariant under the action of the considered subgroup $SO(2)$. 

Continuation to the Minkowski space with the signature $(+,-)$ is carried out via the group replacement $SO(2) \rightarrow SO(1,1)$ -- Lorentz transformations, which can be denoted as $x^2=ix^0$ at the level of spacetime coordinates (here we follow the same notations as in \cite{Alfimov:2020jpy}). Then the spacetime derivatives are changed as follows
\begin{equation}
\partial \rightarrow -\partial_-\,, \quad \bar\partial \rightarrow \partial_+\,,
\end{equation}
whereas the replacement of the fields we define to be
\begin{align}
    & \beta \rightarrow -\beta_-\,, \quad \bar{\beta} \rightarrow \beta_+\,, \quad \gamma\bar{\gamma} \rightarrow -\gamma\bar{\gamma}\,. 
\end{align}
Then the Lagrangian in Minkowski space takes in the form 
\begin{align}\label{Minkowski_betagamma_Lagrangian}
    \mathcal{L}_{\beta\gamma}^{(M)}=\beta_+\partial_-\bar{\gamma}+\beta_-\partial_+\gamma-m \beta_+\beta_--m\bar{\gamma}\gamma+g^2 \beta_+\beta_-\bar{\gamma}\gamma\,.
\end{align}
Therefore, in the Minkowski signature under the Lorentz boosts from $SO(1,1)$ the fields and derivatives transform as follows
\begin{equation}
    \partial_{\pm} \rightarrow e^{\pm \theta}\partial_{\pm}\,, \quad \beta_{\pm} \rightarrow e^{\mp\theta}\beta_{\pm}\,, \quad \gamma \rightarrow \gamma, \quad \bar\gamma\rightarrow \bar\gamma\,,
\end{equation}
where $\theta$ is the rapidity of the boost. Thus we have obtained the formulation of the bosonic Thirring model in the Minkowski signature \eqref{Minkowski_betagamma_Lagrangian}. It is important to notice that the $SO(2)$ subgroup in the symmetry group of the Euclidean version of the model we consider can be chosen in another way, what possibly can lead to another analytic continuation of \eqref{Euclidean_betagamma_Lagrangian} to Minkowski signature.

In the next part we will present another way to derive the action \eqref{Minkowski_betagamma_Lagrangian}.

\subsection{$\beta\gamma$-system from gauged Wess-Zumino-Witten model}

It is interesting that the $\beta\gamma$-system \eqref{Minkowski_betagamma_Lagrangian} can be obtained from the gauged Wess-Zumino-Witten (gWZW) model with integrable potential by certain field redefinitions. In \cite{Hoare:2010fb} there was presented the $SU(2)/U(1)$ gWZW formulation of the CSG model. Namely, its action in the axially gauged case was obtained in the section 3.3.2 of \cite{Hoare:2010fb} and reads
\begin{multline}\label{gauge_fixed_WZW}
    \mathcal{S}_{\textrm{gWZW}}=\frac{k}{4\pi}\int d^2 x \left[\partial_+\phi \partial_-\phi+\sin^2 \phi \partial_+\chi \partial_-\chi- \right. \\
    \left. -a_+\sin^2\phi \partial_-\chi-a_-\sin^2\phi \partial_+\chi-a_+ a_- \cos^2 \phi-\frac{m^2}{2}(\cos 2\phi-1)\right]\,.
\end{multline}
By performing the following shifts of the fields $a_{\pm}$
\begin{equation}
    a_{\pm}=\tilde{a}_{\pm}-\tan^2\phi \partial_{\pm}\chi\,,
\end{equation}
we obtain the transformed action \eqref{gauge_fixed_WZW} in terms of the new fields $\tilde{a}_{\pm}$
\begin{equation}
    \mathcal{S}_{\textrm{gWZW}}=\frac{k}{4\pi}\int d^2 x \left[\partial_+\phi \partial_-\phi+\tan^2 \phi \partial_+\chi \partial_-\chi-\tilde{a}_+ \tilde{a}_- \cos^2 \phi-\frac{m^2}{2}(\cos 2\phi-1)\right]\,.
\end{equation}
The next shift of the fields $\tilde{a}_{\pm}$ would be
\begin{equation}
\tilde{a}_{\pm}=a'_{\pm}-\frac{\partial_{\pm}\left(e^{i\chi}\sin\phi\right)}{\cos^2 \phi}
\end{equation}
and leads us to the action given by the formula (we also made some changes with the field $\phi$ using trigonometric identities)
\begin{multline}
    \mathcal{S}_{\textrm{gWZW}}=\frac{k}{4\pi}\int d^2 x \left[i(\partial_-(\log\cos\phi)\partial_+\chi-\partial_+(\log\cos\phi)\partial_-\chi)+\right. \\
    \left.+a'_+ \partial_-\left(e^{-i\chi}\sin\phi\right)+a'_- \partial_+\left(e^{i\chi}\sin\phi\right)-a'_+ a'_--m^2\sin^2\phi+a'_+ a'_- \sin^2 \phi\right]\,,
\end{multline}
where the integral in the first line of the Lagrangian equals to $0$ for the same reason as in the formulas \eqref{Im_to_polar}, \eqref{f(r)_definition} and \eqref{integration_Im_by_parts} of the \ref{app_Path_integral}. And performing at last the identification of the fields
\begin{equation}\label{field_definitions}
    e^{i\chi}\sin\phi=\sqrt{\frac{4\pi}{mk}}\gamma\,, \quad a'_{\pm}=\sqrt{\frac{4\pi m}{k}}\beta_{\pm}\,,
\end{equation}
we arrive to the action
\begin{equation}\label{betagamma_action_gWZW}
    \mathcal{S}_{\textrm{gWZW}}=\int d^2 x \left[\beta_+\partial_-\bar{\gamma}+\beta_-\partial_+\gamma-m\beta_+\beta_--m\gamma\bar{\gamma}+\frac{4\pi}{k}\beta_+\beta_-\gamma\bar{\gamma}\right]\,,
\end{equation}
whose Lagrangian exactly coincides with \eqref{Minkowski_betagamma_Lagrangian} upon identification of the coupling constant as
\begin{equation}\label{coupling_gWZW}
g^2=\frac{4\pi}{k}\,.
\end{equation}

The next important step is to check that integrating out the fields $\beta_{\pm}$, we are able to reproduce the CSG action with the correct counterterms. Apart from excluding the $\beta_{\pm}$ fields from the action at the classical level by imposing the equations of motion for them, we have to add the functional determinant contribution similar to the one in the \ref{app_Path_integral} $-\log\det(1-g^2 \bar{\gamma}\gamma)$ (up to the $\gamma$ field normalization). 

Following the steps from \cite{Hoare:2010fb} and utilizing the result of \cite{Tseytlin:1991ht}, we obtain a local two-derivative correction coming from the functional determinant
\begin{equation}
    -\frac{1}{8\pi}\left(\frac{4\pi}{k}\right)^2 \frac{\partial_+(\gamma\bar{\gamma})\partial_-(\gamma\bar{\gamma})}{\left(m-\frac{4\pi}{k}\gamma\bar{\gamma}\right)^2}\,,
\end{equation}
which leads to the corrected action of the CSG model
\begin{equation}\label{CSG_from_gWZW}
    \mathcal{S}_{CsG}^{\textrm{corr}}=\int d^2 x \left[\frac{\partial_+\gamma\partial_-\bar{\gamma}}{m-\frac{4\pi}{k}\gamma\bar{\gamma}}-m\gamma\bar{\gamma}-\frac{1}{8\pi}\left(\frac{4\pi}{k}\right)^2 \frac{\partial_+(\gamma\bar{\gamma})\partial_-(\gamma\bar{\gamma})}{\left(m-\frac{4\pi}{k}\gamma\bar{\gamma}\right)^2}\right]\,.
\end{equation}
Then, substituting the $\gamma$ field definition from \eqref{field_definitions} into \eqref{CSG_from_gWZW}, we obtain the following action
\begin{equation}\label{CSG_from_gWZW2}
    \mathcal{S}_{CsG}^{\textrm{corr}}=\int d^2 x \frac{k}{4\pi}\left[\partial_+\phi\partial_-\phi+\tan^2\phi\,\partial_+\chi\partial_-\chi-m^2\sin^2\phi-\frac{2}{k}\tan^2\phi\,\partial_+\phi\partial_-\phi\right]\,,
\end{equation}
exactly coinciding with the action (3.38) from \cite{Hoare:2010fb}. There it was shown that after field rescaling and redefinition one derives the action, which correctly reproduces the S-matrix \eqref{eq:Seaxt1} up to the one-loop order. Summing up, we conclude that the $SU(2)/U(1)$ gWZW model and the $\beta\gamma$-system we consider are related by the field transformation.

In the next section we are going to write the set of Feynman rules for the theory \eqref{Minkowski_betagamma_Lagrangian}, which are subsequently needed for the calculation of the amplitudes.

\subsection{Free field propagators and Feynman rules}

In this part we are going to derive one of several ingredients necessary to calculate the amplitudes, namely, the free field propagators. To find the free field propagators we consider the free classical Lagrangian of $\beta\gamma$-system
\begin{equation}\label{Minkowski_betagamma_Lagrangian_free}
    \mathcal{L}_{\beta\gamma}^{(\textrm{free})}=\beta_+\partial_-\bar{\gamma}+\beta_-\partial_+\gamma-m\beta_+\beta_--m \bar{\gamma}\gamma\,.
\end{equation}
In the matrix form the free Lagrangian \eqref{Minkowski_betagamma_Lagrangian_free} can be rewritten as follows
\begin{equation}
    \mathcal{L}_{\beta\gamma}^{(\textrm{free})}=-\frac{1}{2}
    \begin{pmatrix}
    \beta_+ & \beta_- & \bar{\gamma} & \gamma
    \end{pmatrix}
    \underbrace{\begin{pmatrix}
    0 & m & -\partial_- & 0 \\
    m & 0 & 0 & -\partial_+ \\
    \partial_- & 0 & 0 & m \\
     0 & \partial_+ & m & 0
    \end{pmatrix} }_{\hat{K}}
    \begin{pmatrix}
    \beta_+\\
    \beta_-\\
    \bar{\gamma}\\
    \gamma
    \end{pmatrix}\,.
\end{equation}
Let us apply the Fourier transform to the free action
\begin{equation}\label{Free_betagamma_action_Fourier}
    S_{\beta\gamma}^{(\textrm{free})}=-\frac{1}{2}\int\frac{d^2 k}{(2\pi)^2}
    \begin{pmatrix}
    \tilde{\beta}_+(k) \, \tilde{\beta}_-(k) \, \tilde{\bar{\gamma}}(k) \, \tilde{\gamma}(k)
    \end{pmatrix}
    \underbrace{\begin{pmatrix}
    0 & m & -2ik^+ & 0 \\
    m & 0 & 0 & -2ik^- \\
    2ik^+ & 0 & 0 & m \\
    0 & 2ik^- & m & 0
    \end{pmatrix} }_{\tilde{\hat{K}}(k)}
    \begin{pmatrix}
    \tilde{\beta}_+(-k) \\
    \tilde{\beta}_-(-k) \\
    \tilde{\bar{\gamma}}(-k) \\
    \tilde{\gamma}(-k)
    \end{pmatrix}\,,
\end{equation}
where we used the following definitions for the lightcone variables
\begin{equation}
    x^{\pm}=\frac{x^0 \pm x^1}{2}\,, \quad \partial_{\pm}=\partial_0 \pm \partial_1\,, \quad k^{\pm}=\frac{k^0\pm k^1}{2}\,, \quad (k,x) = k^0 x^0 - k^1 x^1 = 2(k^+ x^- + k^- x^+)\,.
\end{equation}

\begin{figure}[h!]\centering
     \def\svgwidth{11cm} 
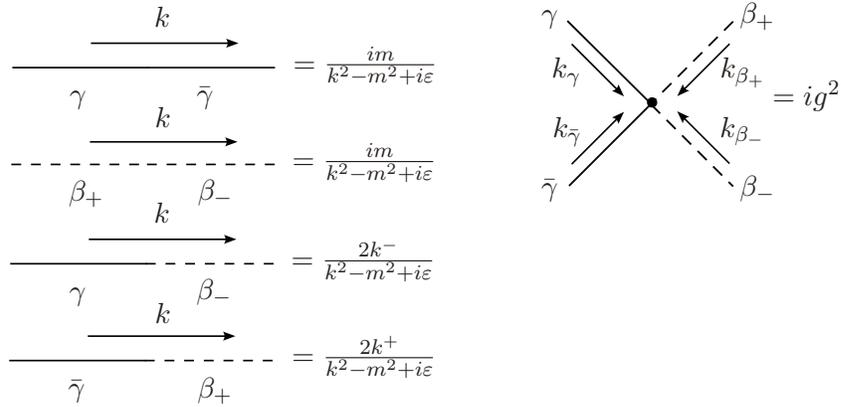   \caption{Feynman rules for the Minkowskian $\beta\gamma$-system \eqref{Minkowski_betagamma_Lagrangian}.}\label{Rules}
\end{figure}

In the formula \eqref{Free_betagamma_action_Fourier} we introduced the kinetic operator $\tilde{\hat{K}}(k)$ in Fourier space. In order to determine the set of propagators of the $\beta\gamma$ theory, we can calculate the inverse matrix of the operator $\tilde{\hat{K}}(k)$ as
\begin{equation}
    \tilde{\hat{K}}^{-1}(k)=\begin{pmatrix}
    0 & \frac{-m}{k^2-m^2} & \frac{-2ik^-}{k^2-m^2} & 0\\
    \frac{-m}{k^2-m^2} & 0 & 0 & \frac{-2ik^+}{k^2-m^2} \\
    \frac{2ik^-}{k^2-m^2} & 0 & 0 & \frac{-m}{k^2-m^2} \\
    0 & \frac{2ik^+}{k^2-m^2} & \frac{-m}{k^2-m^2} & 0
    \end{pmatrix}\,.
\end{equation}
Then the propagator matrix is found as $\hat{D}(k) = -i\hat{\tilde K}^{-1}(k)$. Thus, we have the following set of propagators in the Fourier representation \footnote{In the following sections, the prescription $+i\varepsilon$ is omitted in the expressions in the Green's function expressions.}
\begin{align}
    & \langle\tilde{\beta}_+(k)\tilde{\beta}_-(-k)\rangle =\langle\tilde{\beta}_-(k)\beta_+(-k)\rangle = \langle\tilde{\gamma}(k)\tilde{\bar{\gamma}}(-k)\rangle = \langle\tilde{\bar{\gamma}}(k)\tilde{\gamma}(-k)\rangle=\frac{im}{k^2-m^2+i\varepsilon}\,,
    \\
    & \langle\tilde{\bar{\gamma}}(k)\tilde{\beta}_+(-k)\rangle = -\langle\tilde{\beta}_+(k)\tilde{\bar{\gamma}}(-k)\rangle  =\frac{2k^-}{k^2-m^2+i\varepsilon}\,, \quad
    \langle\tilde{\gamma}(k)\tilde{\beta}_-(-k)\rangle=-\langle \tilde{\beta}_-(k)\tilde{\gamma}(-k)\rangle=\frac{2k^+}{k^2-m^2+i\varepsilon}\,. \notag
\end{align}
Having the free field propagators we are able to sum up the results of the present subsection by writing down the Feynman rules of the theory, which are shown in the Figure~\ref{Rules}.

In the next Section we are going to study the renormalization properties of our theory by exploring the two-point functions up to two loops. This information is essential for our main goal of finding the $2 \rightarrow 2$ amplitudes in the sector of the particles $\gamma$ and $\bar{\gamma}$.

\section{Two-point correlation functions}\label{Two_point_functions}

This part is devoted to the calculation of two-point functions of the Lagrangian fields \eqref{Minkowski_betagamma_Lagrangian} at one and two loops. We have to investigate the divergences appearing in these functions to understand, which counterterms are needed to compensate them. Let us start from the one loop order.

\subsection{One-loop two-point correlation functions}

\begin{figure}[h!]\centering
      \def\svgwidth{16cm} 
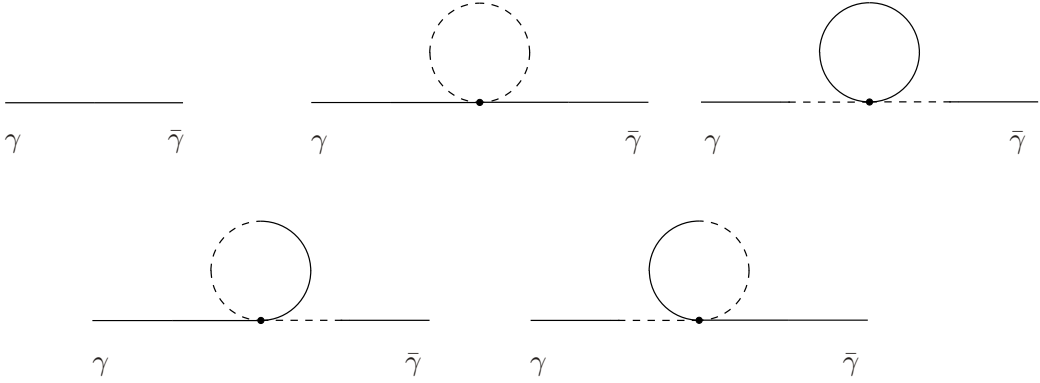\caption{Contribution of one-loop diagrams  into the $\gamma\bar\gamma$ propagator.}

\label{1_loop_2_point_diagrams}
\end{figure}

For the correlation function $\langle\tilde{\gamma}(k)\tilde{\bar{\gamma}}(-k)\rangle$ up to the one loop order only the diagrams shown in the Figure~\ref{1_loop_2_point_diagrams} contribute.
For the other fields the propagators can vary, however, the form of all such diagrams is the same. It should be noted that due to the symmetry of loop integral the diagrams with the half-continuous half-dashed propagator are equal to $0$. Thus, we obtain the following result
\begin{align}\label{1_loop_2_point_functions}
    \langle\tilde{\gamma}(k)\tilde{\bar{\gamma}}(-k)\rangle^{(1)}&=g^2 \frac{m(k^2+m^2)}{(k^2-m^2)^2}\int\frac{d^2 p}{(2\pi)^2}\frac{1}{p^2-m^2}\,, \\
    \langle\tilde{\beta}_+(k)\tilde{\beta}_-(-k)\rangle^{(1)}&=g^2 \frac{m(k^2+m^2)}{(k^2-m^2)^2}\int\frac{d^2 p}{(2\pi)^2}\frac{1}{p^2-m^2}\,, \notag \\
    \langle\tilde{\bar{\gamma}}(k)\tilde{\beta}_+(-k)\rangle^{(1)}&=-ig^2 \frac{4m^2k^+}{(k^2-m^2)^2}\int\frac{d^2 p}{(2\pi)^2}\frac{1}{p^2-m^2}\,, \notag 
    \\
    \langle\tilde{\gamma}(k)\tilde{\beta}_-(-k)\rangle^{(1)}&=-ig^2 \frac{4m^2k^-}{(k^2-m^2)^2}\int\frac{d^2 p}{(2\pi)^2}\frac{1}{p^2-m^2}\,, \notag
\end{align}
where the index $(1)$ denotes the one-loop contribution to the correlation function. We see that all the 1-loop contributions to the two point functions \eqref{1_loop_2_point_functions} are expressed in terms of the one-loop integral only, which can be computed by using the dimensional regularization ($d=2-2\epsilon$). The two-point functions at one-loop level $g^2$ are
\begin{align}\label{1_loop_2_point_functions_divergencies}
    \langle\tilde{\gamma}(k)\tilde{\bar{\gamma}}(-k)\rangle^{(1)}&=-g^2\frac{im(k^2+ m^2)}{4\pi(k^2-m^2)^2}\left(\frac{1}{\epsilon}-\log\frac{m^2}{\mu^2}-\gamma_E+\log(4\pi)+\mathcal{O}(\epsilon)\right)\,, \\
    \langle\tilde{\beta}_+(k)\tilde{\beta}_-(-k)\rangle^{(1)}&=-g^2\frac{im(k^2+ m^2)}{4\pi(k^2-m^2)^2}\left(\frac{1}{\epsilon}-\log\frac{m^2}{\mu^2}-\gamma_E+\log(4\pi)+\mathcal{O}(\epsilon)\right)\,, \notag \\
    \langle\tilde{\bar{\gamma}}(k)\tilde{\beta}_+(-k)\rangle^{(1)}&=-g^2\frac{m^2k^+}{\pi(k^2-m^2)^2}\left(\frac{1}{\epsilon}-\log\frac{m^2}{\mu^2}-\gamma_E+\log(4\pi)+\mathcal{O}(\epsilon)\right)\,, \notag 
    \\
    \langle\tilde{\gamma}(k)\tilde{\beta}_-(-k)\rangle^{(1)}&=-g^2\frac{m^2k^-}{\pi(k^2-m^2)^2}\left(\frac{1}{\epsilon}-\log\frac{m^2}{\mu^2}-\gamma_E+\log(4\pi)+\mathcal{O}(\epsilon)\right)\,, \notag
\end{align}
\begin{figure}[h!]\centering
      \def\svgwidth{7cm} 
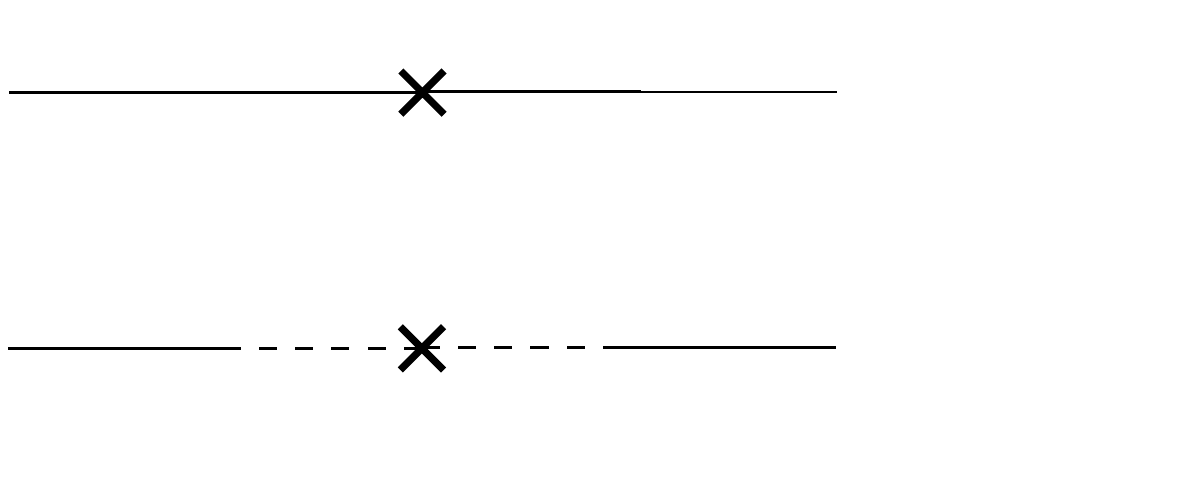

\caption{One-loop diagrams corresponding to the contributions of the counterterms to the $\gamma\bar{\gamma}$ propagators.}
\label{1_loop_2_point_diagrams_conterterms}
\end{figure}
where $\gamma_E$ is the Euler-Mascheroni constant. Now we are able to find the counterterms for the Lagrangian, which compensate the divergencies in \eqref{1_loop_2_point_functions_divergencies}. Our observation is that at one loop it is enough to renormalize the mass terms only
\begin{equation}\label{eq:Zmatrix}
        m \rightarrow Z_{m}m=(1+\delta_m)m=\left(1+g^2 \delta_m^{(1)}+\mathcal{O}\left(g^4\right)\right)m\,,
\end{equation}
where $m$ is  physical mass (we assume that  $m_{\beta} = m_{\gamma} = m$). Then the Lagrangian, up to terms of order $\sim g^2$, takes the following form
\begin{align}\label{Minkowski_betagamma_Larangian_renormalized}
    &\mathcal{L}=\beta_+\partial_-\gamma+\beta_-\partial_+\bar\gamma-m\gamma\bar\gamma-m\beta_+\beta_-+g^2\beta_+\beta_-\gamma\bar\gamma-g^2\delta_m^{(1)} m\beta_+ \beta_--g^2\delta_m^{(1)} m\bar{\gamma}\gamma\,.
\end{align}
Taking into account additional vertices, we get the complementary set of counterterm diagrams, which are drawn in the Figure~\ref{1_loop_2_point_diagrams_conterterms} for the case of $\gamma\bar{\gamma}$ propagator. For the other propagators the diagrams look similar.

Therefore, the contribution of the counterterms equals to
\begin{align}\label{1_loop_2_point_functions_counterterms}
    \langle\tilde{\gamma}(k)\tilde{\bar{\gamma}}(-k)\rangle^{(1)}&=ig^2 \frac{m(k^2+m^2)}{(k^2-m^2)^2}\delta_m^{(1)}\,, \\
    \langle\tilde{\beta}_+(k)\tilde{\beta}_-(-k)\rangle^{(1)}&=ig^2 \frac{m(k^2+m^2)}{(k^2-m^2)^2}\delta_m^{(1)}\,, \notag \\
    \langle\tilde{\bar{\gamma}}(k)\tilde{\beta}_+(-k)\rangle^{(1)}&=g^2 \frac{4m^2k^+}{(k^2-m^2)^2}\delta_m^{(1)}\,, \notag \\
    \langle\tilde{\gamma}(k)\tilde{\beta}_-(-k)\rangle^{(1)}&=g^2 \frac{4m^2k^-}{(k^2-m^2)^2}\delta_m^{(1)}\,. \notag
\end{align}
Combining the diagram contributions corresponding to \eqref{1_loop_2_point_functions} together with the counterterm contributions \eqref{1_loop_2_point_functions_counterterms}, we find the following condition for the counterterm coefficient $\delta_m^{(1)}$
\begin{equation}\label{deltam1_counterterm}
    \delta_m^{(1)} =\frac{\Gamma(\epsilon)}{(4\pi)^{1-\epsilon}}\left(\frac{\mu}{m}\right)^{2\epsilon} = \frac{1}{4\pi}\left(\frac{1}{\epsilon}-\log\frac{m^2}{\mu^2}-\gamma_E +\log(4\pi)+ \mathcal{O}(\epsilon)\right)\,.
\end{equation}

Now let us turn to the analysis of the two loop contributions to the two-point functions.

\subsection{Two-loop two-point functions}

\begin{figure}[h!]\centering
      \def\svgwidth{16cm} 
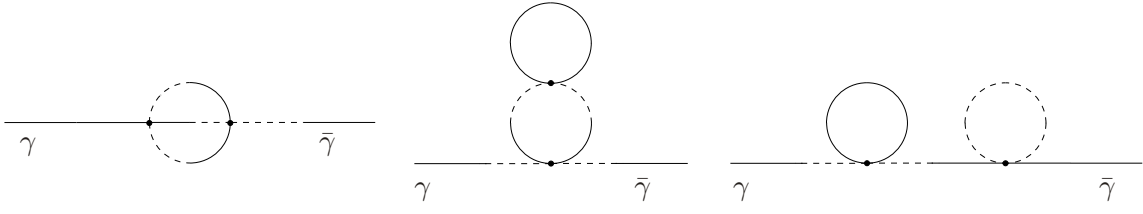    \caption{Two-loop diagrams for the $\gamma\bar{\gamma}$ propagator.}

\label{2_loop_2_point_diagrams}
\end{figure}

At the two-loop level ($\sim g^4$) the two-point function is represented by sunset diagrams, diagrams with two connected loops and two loops separated by a propagator. The corresponding diagrams for the case of $\gamma\bar{\gamma}$ propagator are shown on the Figure~\ref{2_loop_2_point_diagrams}. These types of diagrams also contain divergent terms. Let us denote the contributions coming from these three types by a, b and c correspondingly. We are going to consider the example of the $\gamma\bar{\gamma}$ propagator.

Let us start from writing down the sunset diagram contributions. We can classify them by the number of half-dashed half-continuous propagators in the loops, therefore there are 4 types of of loop integrals in the sunset diagrams
\begin{align}\label{sunset_loop_integrals}
    J_0(k^2) &= -\int\frac{d^2p_1}{(2\pi)^2}\frac{d^2p_2}{(2\pi)^2}\,\frac{im^3}{(p_1^2-m^2+i\varepsilon)(p_2^2-m^2+i\varepsilon)((k-p_1-p_2)^2-m^2+i\varepsilon)}\,,
    \\
    J_1^{\pm}(k) &=-\int\frac{d^2p_1}{(2\pi)^2}\frac{d^2p_2}{(2\pi)^2}\,\frac{2m^2p_1^{\pm}}{(p_1^2-m^2+i\varepsilon)(p_2^2-m^2+i\varepsilon)((k-p_1-p_2)^2-m^2+i\varepsilon)}\,, \notag
    \\
    J_2^{\pm\mp}(k^2) &= \int\frac{d^2p_1}{(2\pi)^2}\frac{d^2p_2}{(2\pi)^2}\,\frac{4imp_1^{\pm}p_2^{\mp}}{(p_1^2-m^2+i\varepsilon)(p_2^2-m^2+i\varepsilon)((k-p_1-p_2)^2-m^2+i\varepsilon)}\,, \notag
    \\
    J_3^{\pm\mp\pm}(k) &= \int\frac{d^2p_1}{(2\pi)^2}\frac{d^2p_2}{(2\pi)^2}\,\frac{8p_1^{\pm}p_2^{\pm}(k-p_1-p_2)^{\mp}}{(p_1^2-m^2+i\varepsilon)(p_2^2-m^2+i\varepsilon)((k-p_1-p_2)^2-m^2+i\varepsilon)}\,, \notag
\end{align}
where the one depicted in the Figure~\ref{2_loop_2_point_diagrams} corresponds to the $J_3$ type. By combining all the diagrams of the first type we obtain the total sunset diagram contribution to the two-point function $\gamma\bar{\gamma}$
\begin{align}\label{group_a_diagrams}
    &\langle \tilde\gamma(k)\tilde{\bar\gamma}(-k)\rangle_a^{(2)} = g^4\frac{k^2+m^2}{(k^2-m^2)^2}J_0(k^2)-\frac{2img^4}{(k^2-m^2)^2}(k^+J^-_1(k)+k^-J_1^+(k)) -
    \\
    &-g^4\frac{k^2+m^2}{(k^2-m^2)^2}J_2^{+-}(k^2)+\frac{2img^4}{(k^2-m^2)^2}(k^+J_3^{-+-}(k)+k^-J_3^{+-+}(k)) =\notag
    \\
    &=g^4\frac{k^2+m^2}{(k^2-m^2)^2}\left(J_0(k^2)-J_2^{+-}(k^2)\right)-\frac{4img^4}{(k^2-m^2)^2}k^+\left(J^-_1(k)-k^+J_3^{-+-}(k)\right)
    \,. \notag 
\end{align}
The sunset loop integrals can be expressed one in terms of the other, which is partially done in the~\ref{J_integrals_indetities}.

The next contribution comes from the diagrams with two connected loops of the type b, which give the result
\begin{align}\label{group_b_diagrams}
    &\langle \tilde\gamma(k)\tilde{\bar\gamma}(-k)\rangle_b^{(2)} = g^4\frac{k^2+m^2}{(k^2-m^2)^2}(I_1(0)+I_2^{+-}(0))\int\frac{d^2p}{(2\pi)^2}\frac{im}{p^2-m^2}= \\
    & =\frac{ig^4m(k^2+m^2)}{16\pi^2(k^2-m^2)^2}\left(\frac{1}{\epsilon}-\log\frac{m^2}{\mu^2}-\gamma_E+\log(4\pi)+ \mathcal{O}(\epsilon)\right)\left(\frac{1}{\epsilon}-2-\gamma_E+\log(4\pi) - \log\frac{m^2}{\mu^2}+ \mathcal{O}(\epsilon)\right) \,, \notag
\end{align}
where the loop integrals $I_1(k^2)$ \eqref{1st_group_loop_integrals}, $I_2^{+-}(k^2)$ \eqref{2nd_group_loop_integrals} and $I_3^{\pm}(k^2)$ \eqref{3rd_group_loop_integrals} are defined in the Section~\ref{1_loop_4_point_function} and calculated in the~\ref{app_I1_integral}, \ref{app_I2_integral} and \ref{app_I3_integral}.

And the last is the contribution from the third group c of diagrams, for which the loops are separated by a propagator
\begin{align}\label{group_c_diagrams}
    & \langle\tilde\gamma(k)\tilde{\bar\gamma}(-k)\rangle_c^{(2)} = g^4\frac{im(3k^2+m^2)}{(k^2-m^2)^3}\int\frac{d^2p}{(2\pi)^2}\frac{im}{p^2-m^2}\int\frac{d^2p'}{(2\pi)^2}\frac{im}{p'^2-m^2}=
    \\
    &=\frac{ig^4m^3(3k^2+m^2)}{16\pi^2(k^2-m^2)^3}\left(\frac{1}{\epsilon}-\log\frac{m^2}{\mu^2}-\gamma_E+\log(4\pi)+ \mathcal{O}(\epsilon)\right)^2\,.\notag
\end{align}


To see which counterterms are needed at the two loop order let us write down the contribution of diagrams containing the $\delta_m^{(1)}$ counterterm vertices into the two-point function up to the order $\sim g^4$. This contribution is given by
\begin{align}\label{2_loop_gammagammabar_counterterms}
    & \langle\tilde\gamma(k)\tilde{\bar\gamma}(-k)\rangle_{\delta_m^{(1)}-\textrm{counterterms}}^{(2)}=-g^4\frac{k^2+m^2}{(k^2-m^2)^2}(I_1(0)+I_2^{+-}(0))\delta_m^{(1)}m-
    \\
    & -2g^4\frac{2im\left(3k^2+m^2\right)}{(k^2-m^2)^3}m\delta_m^{(1)}\int \frac{d^2p}{(2\pi)^2}\frac{im}{p^2-m^2}+g^4\frac{im\left(3k^2+m^2\right)}{(k^2-m^2)^3}(m\delta_m^{(1)})^2 = \notag
    \\
    &= -\frac{ig^4m^3\left(3k^2+ m^2\right)}{16\pi^2(k^2-m^2)^3}\left(\frac{1}{\epsilon}-\log\frac{m^2}{\mu^2}-\gamma_E+\log(4\pi)+ \mathcal{O}(\epsilon)\right)^2-\notag
    \\
    &-\frac{ig^4m\left(m^2+k^2\right)}{16\pi^2(k^2-m^2)^2}\left(\frac{1}{\epsilon}-2-\gamma_E+\log(4\pi) - \log\frac{m^2}{\mu^2}+ \mathcal{O}(\epsilon)\right)\left(\frac{1}{\epsilon}-\log\frac{m^2}{\mu^2}-\gamma_E+\log(4\pi)+ \mathcal{O}(\epsilon)\right)\,.\notag
\end{align}
If we compare \eqref{2_loop_gammagammabar_counterterms} with \eqref{group_b_diagrams} and \eqref{group_c_diagrams} we see that the diagrams from these groups are exactly compensated by the corresponding counterterms from \eqref{2_loop_gammagammabar_counterterms} containing $\delta_m^{(1)}$. It means that the only remaining divergence comes from the sunset diagrams contribution to the $\gamma\bar{\gamma}$ correlator
\begin{align}
    &\langle \tilde\gamma(k)\tilde{\bar\gamma}(-k)\rangle_a^{(2)} = \frac{img^4}{16\pi^2}\frac{k^2+m^2}{(k^2-m^2)^2}\left(\frac{1}{\epsilon^2} + \frac{2}{\epsilon}\bigg(-\gamma_E+\log4\pi - \log\left(\frac{m}{\mu}\right)^2\right) +\mathcal{O}(\epsilon^0) \,.
\end{align}
Then, in the framework of the minimal subtraction scheme, we can write the counterterm coefficient $\delta_m^{(2)}$ in the form
\begin{equation}\label{deltam2_conterterm}
   \delta_m^{(2)}= \frac{1}{16\pi^2}\left(\frac{1}{\epsilon^2}+\frac{2}{\epsilon}\left(-\gamma_E + \log4\pi -\log\left(\frac{m}{\mu}\right)^2\right)\right)\,.
\end{equation}

In addition, one can calculate the other two-point functions involving the fields $\beta$. All the diagrams except for the sunset ones, are exactly annihilated by the counterterm diagrams with $\delta_m^{(1)}$ \eqref{deltam1_counterterm}. Whereas the sunset diagram contributions are given by the following result
\begin{align}\label{2_loop_2_point_beta}
    \langle\tilde{\beta}_+(k)\tilde{\beta}_-(-k)\rangle_a^{(2)}&=g^4\frac{k^2+m^2}{(k^2-m^2)^2}\left(J_0(k^2)-J_2^{+-}(k^2)\right)-\frac{4img^4}{(k^2-m^2)^2}k^+\left(J^-_1(k)-J_3^{-+-}(k)\right)\,, \notag
    \\
    \langle\tilde{\gamma}(k)\tilde{\beta}_-(-k)\rangle_a^{(2)}&=\frac{4g^4imk^-}{(k^2-m^2)^2}(J_2^{+-}(k^2)-J_0(k^2))+\frac{4g^4(k^-)^2}{(k^2-m^2)^2}(J_3^{++-}(k)-J_1^+(k))+
    \\&+\frac{g^4m^2}{(k^2-m^2)^2}(J_3^{+--}(k)-J_1^-(k))\,, \notag
    \\
    \langle\tilde{\bar\gamma}(k)\tilde{\beta}_+(-k)\rangle_a^{(2)}&=\frac{4g^4imk^+}{(k^2-m^2)^2}(J_2^{+-}(k^2)-J_0(k^2))+\frac{4g^4(k^+)^2}{(k^2-m^2)^2}(J_3^{+--}(k)-J_1^-(k))+ \notag
    \\
    &+\frac{g^4m^2}{(k^2-m^2)^2}(J_3^{++-}(k)-J_1^+(k))\,. \notag
\end{align}
It is not complicated to check that the counterterm diagrams with $\delta_m^{(2)}$ \eqref{deltam2_conterterm} compensate the divergences in \eqref{2_loop_2_point_beta} as well.

In the present section we showed that at one loop the divergences in the two-point function can be compensated by the renormalization of the mass only and we do not need any renormalization of the fields at one loop. At two loops we found that the contributions at this order proportional to $\delta_m^{(1)}$ compensate the contributions of all the diagrams into the $\gamma\bar{\gamma}$ correlation function except for the sunset ones. For these diagrams we showed that the divergencies can be absorbed into $\delta_m^{(2)}$.

In the next Section we are going to calculate the $2 \rightarrow 2$ amplitudes in the sector of the particles $\gamma$ and $\bar{\gamma}$ by extracting them from the four-point correlation function with the usage of the Lehman-Symanczik-Zimmermann (LSZ) reduction formula in order to compare them with the $S$-matrix results from \cite{deVega:1981ka} and \cite{Dorey:1994mg}.

\section{Four-point correlation functions}\label{Four_point_functions}

The main objective of our paper is to compare the amplitudes coming from the Lagrangian \eqref{Minkowski_betagamma_Larangian_renormalized} with the ones calculated in \cite{deVega:1981ka} and \cite{Dorey:1994mg}. We are concentrated on the $2 \rightarrow 2$ amplitudes corresponding to the fields $\gamma$
and $\bar{\gamma}$. To achieve this goal one can use the correlation functions
\begin{equation}
\langle\gamma(x_1)\bar{\gamma}(x_2)\gamma(x_3)\bar{\gamma}(x_4)\rangle\,, \quad \langle\gamma(x_1)\gamma(x_2)\gamma(x_3)\gamma(x_4)\rangle\,, \quad \langle\bar{\gamma}(x_1)\bar{\gamma}(x_2)\bar{\gamma}(x_3)\bar{\gamma}(x_4)\rangle\,.
\end{equation}
Given the Feynman rules from the Figure~\ref{Rules} it is more convenient to calculate the corresponding correlation functions in the Fourier representation
\begin{equation}\label{4_point_correlators}
\langle\tilde{\gamma}_1\tilde{\bar{\gamma}}_2\tilde{\gamma}_3\tilde{\bar{\gamma}}_4\rangle\,, \quad \langle\tilde{\gamma}_1\tilde{\gamma}_2\tilde{\gamma}_3\tilde{\gamma}_4\rangle\,, \quad \langle\tilde{\bar{\gamma}}_1\tilde{\bar{\gamma}}_2\tilde{\bar{\gamma}}_3\tilde{\bar{\gamma}}_4\rangle\,,
\end{equation}
where $\tilde{\gamma}_i \overset{\textrm{def}}{=} \tilde{\gamma}(k_i)$ for brevity. As we already mentioned in the Section~\ref{CsG_setting} the particle $\gamma$ is associated to the soliton with the charge $Q=1$ and the particle $\bar{\gamma}$ corresponds to the soliton with the charge $Q=-1$ in the CSG theory and, in addition, all other $2 \rightarrow 2$ amplitudes like $\gamma\gamma \rightarrow \gamma\gamma$ and $\bar{\gamma}\bar{\gamma} \rightarrow \bar{\gamma}\bar{\gamma}$ are expressed in terms of the only one function \eqref{eq:Seaxt1}. Therefore, it is sufficient to study the following correlation function
\begin{equation}
\langle\tilde{\gamma}_1\tilde{\bar{\gamma}}_2\tilde{\gamma}_3\tilde{\bar{\gamma}}_4\rangle=\sum\limits_{n=1}^{+\infty}\tilde{G}^{(n)}(k_1,k_2,k_3,k_4)
\end{equation}
as the others are related to it by the crossing symmetry transformation ($\tilde{G}^{(k)}(k_1,k_2,k_3,k_4)$ corresponds to the $k$-th loop contribution). In the following subsections we will calculate the four point correlation functions in question at the tree and one-loop level.

\subsection{Tree-level four-point function}

\begin{figure}[h!]\centering
\def\svgwidth{7.5cm} 
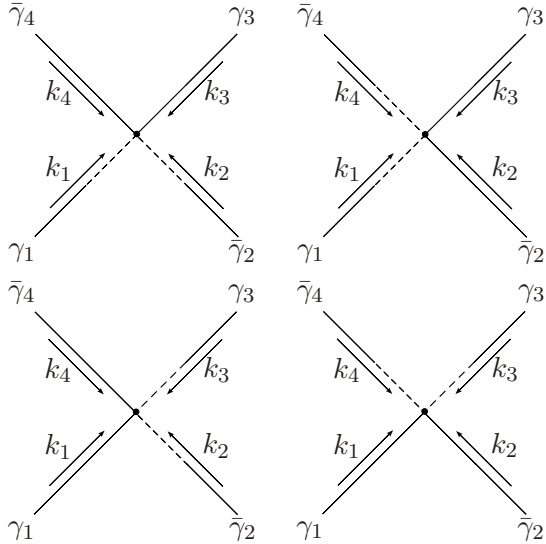
\caption{Tree level Feynman diagrams for the correlation function $\langle\tilde{\gamma}_1\tilde{\bar{\gamma}}_2\tilde{\gamma}_3\tilde{\bar{\gamma}}_4\rangle$.}
\label{tree_level_4_point_diagrams}
\end{figure}

According to our Feynman rules, at the tree level the four-point correlation function $\langle\tilde{\gamma}_1\tilde{\bar{\gamma}}_2\tilde{\gamma}_3\tilde{\bar{\gamma}}_4\rangle$ is given by following set of diagrams in the Figure~\ref{tree_level_4_point_diagrams}, which correspond to the connected diagrams at the order $g^2$.
Using the Feynman rules defined in the Section~\ref{betagamma_system_continuation}, we write the four-point function in momentum space as the product of vertex term and four propagators connecting the vertex with the external lines
\begin{equation}
    \tilde{G}^{(1)}(k_1,k_2,k_3,k_4)=-g^2\frac{4im^2 (k_1^-+k_3^-)(k_2^++k_4^+)}{(k_1^2-m^2)(k_2^2-m^2)(k_3^2-m^2)(k_4^2-m^2)}\,,
\end{equation}
where the external momenta are directed towards the vertex. It should be noted that for the other two correlation functions from \eqref{4_point_correlators} on the tree level we have
\begin{equation}
    \langle\gamma_1\gamma_2\gamma_3\gamma_4\rangle=\mathcal{O}(g^4)\,, \quad \langle\bar\gamma_1\bar\gamma_2\bar\gamma_3\bar\gamma_4\rangle=\mathcal{O}(g^4)\,.
\end{equation}

The tree level amplitude can be obtained by applying the LSZ reduction formula to the four-point function. As the result the tree-level amplitude takes the following form
\begin{equation}\label{eq:matrixTree1}
    \mathcal{M}_{\gamma\bar{\gamma}}^{\gamma\bar{\gamma}(tree)}=-4ig^2 m^2 (k_1^-+k_3^-)(k_2^++k_4^+)=
    \begin{cases} 
    & -4ig^2 m^4 \sinh^2\frac{\theta_1 - \theta_2}{2}\,,\quad k_3=-k_2\,, \; k_4=-k_1\,, \\
    & 0\,, \quad k_3=-k_1\,, \; k_4=-k_2\,,
\end{cases} 
\end{equation}
where $\theta_1,\,\theta_2$ are rapidities of $\gamma_1$ and $\bar\gamma_2$ respectively. We also took into account that the energy-momentum conservation laws for the $(1+1)$-dimensional case has only two solutions
\begin{equation}
k_3=-k_2\,, \;k_4=-k_1\; \text{(transition)}\quad \text{or} \quad k_3=-k_1\,, \;k_4=-k_2\;\text{(reflection)}.
\end{equation}
It can be seen that the requirement of the reflectionless of the scattering at the tree level is satisfied, and, therefore, the Yang-Baxter equation is fulfilled. If we pass to the dimensionless fields $\gamma\,, \bar{\gamma} \rightarrow \gamma/\sqrt{m}\,, \bar{\gamma}/\sqrt{m}$ then the tree-level amplitude of $\gamma\bar{\gamma} \rightarrow \gamma\bar{\gamma}$ scattering has the form
\begin{align}\label{tree_level_gammagammabar_amplitude}
    \mathcal{M}_{\gamma\bar{\gamma}}^{\gamma\bar{\gamma}(tree)}(\theta)= -4ig^2 m^2 \sinh^2\frac{\theta}{2}\,,
\end{align}
where we use the notation $\theta=\theta_1-\theta_2$.

For the special case of $2 \rightarrow 2$ scattering, the $S$-matrix element is related to the amplitude by the following formula
\begin{equation}\label{2_particle_S_matrix}
    S(\theta) = 1+\frac{\mathcal{M}(\theta)}{4m^2\sinh\theta}\,.
\end{equation}
Therefore, combining \eqref{2_particle_S_matrix} with \eqref{tree_level_gammagammabar_amplitude}, we obtain the $\gamma\bar{\gamma} \rightarrow \gamma\bar{\gamma}$ scattering amplitude $S_{\gamma\bar{\gamma}}^{\gamma\bar{\gamma}}(\theta)$ at the tree level takes the following form
\begin{align}
    & S_{\gamma\bar{\gamma}}^{\gamma\bar{\gamma}}(\theta)=1-\frac{ig^2}{2}\left(\coth\theta-\cosech\theta\right) +\mathcal{O}(g^4)=1-\frac{ig^2}{2}\tanh\frac{\theta}{2}+\mathcal{O}(g^4)\,,
\end{align}
which coincides with the CSG results \eqref{eq:S-matrix} and \eqref{eq:Seaxt2} at the tree level.

To sum up, in the subsection above we managed show that the tree level amplitudes for the elementary excitations generated by the fields $\gamma$ and $\bar{\gamma}$ obtained from the Lagrangian \eqref{Minkowski_betagamma_Larangian_renormalized} exactly coincide with the same tree level amplitudes from \cite{deVega:1981ka} and \cite{Dorey:1994mg}. In the subsequent subsection we will explore such a relation at the one loop level.

\subsection{One-loop four-point function}\label{1_loop_4_point_function}

\begin{figure}[h!]\centering
     \def\svgwidth{17cm} 
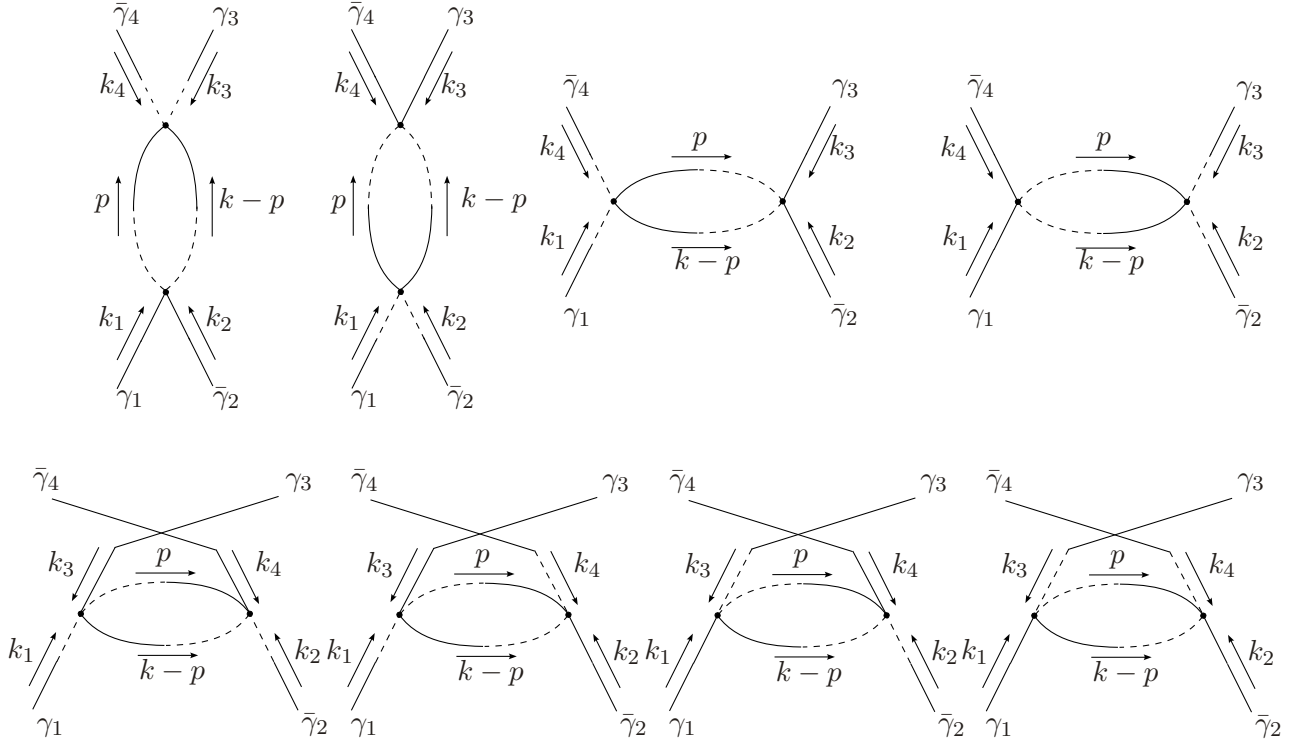
\caption{One-loop contributions with the $+-$ Lorentz structure in the loop diagrams.}
\label{1_loop_4_point_2nd_type}
\end{figure}

Let us consider the four-point correlation function $\langle \tilde\gamma_1 \tilde{\bar\gamma}_2 \tilde\gamma_3 \tilde{\bar\gamma}_4 \rangle$ at the one-loop level. All the connected diagrams giving nonzero contribution at this order have the structure of a loop consisting of two propagators with two pairs of external lines connected to it. All one-loop diagrams can be split into 3 groups according to the set of propagators forming the loops. They are: with continuous or dashed propagators only, with two half-dashed half-continuous propagators and with one half-dashed half-continuous propagator. As an illustration we drew all the diagrams of the second type, which have the Lorentz indices $+-$ in the loop integral in the Figure~\ref{1_loop_4_point_2nd_type}.

The first group are the diagrams compiled by contractions of the same type fields, i.e. the contractions of the types $\wick{\c1\beta_+(x)\c1\beta_-(y)}$, $\wick{\c1\gamma(x)\c1{\bar\gamma}(y)}$. The corresponding loop integral takes the form   
\begin{equation}\label{1st_group_loop_integrals}
        I_1\left(k^2\right)=-\int\frac{d^2 p}{(2\pi)^2}\frac{m^2}{(p^2-m^2+i\varepsilon)((k-p)^2-m^2+i\varepsilon)}\,.
\end{equation}
As we have the $s$-, $t$- and $u$- channel diagrams, the loop integral of the first group $I_1$ results in (details can be found in the~\ref{app_loop_integrals})
\begin{align}
    & I_1\left(s=(k_1+k_2)^2\right) =\frac{i}{4\pi}\frac{\theta-i\pi}{\sinh\theta}\,, \\
    & I_1\left(t=(k_1+k_4)^2\right)=
    \begin{cases}
    -\frac{i}{4\pi}\frac{\theta }{\sinh\theta}\,, \quad k_3=-k_1\,, \quad k_4 = -k_2\,, \\
    -\frac{i}{4\pi}\,, \quad k_4 = -k_1,\; k_3=-k_2\,,
    \end{cases} \notag
    \\
    & I_1\left(u=(k_1+k_3)^2\right)=
    \begin{cases}
    -\frac{i}{4\pi}\,, \quad k_3 = -k_1\,, \quad k_4= -k_2\,, \\
    -\frac{i}{4\pi}\frac{\theta }{\sinh\theta},\; k_4=-k_1,\, k_3=-k_2\,.
    \end{cases} \notag  
\end{align}

The second group are the diagrams originating from the contractions of the different type fields, i.e. from the contractions of the type
$\wick{\c1\beta_+(x)\c1\gamma(y)},\,\wick{\c1\beta_-(x)\c1{\bar\gamma}(y)},\,\wick{\c1\gamma(x)\c1\beta_+(y)},\,\wick{\c1{\bar\gamma}(x)\c1\beta_-(y)}$. There are two expressions appearing in the loop integral
\begin{align}\label{2nd_group_loop_integrals}
        & I_2^{\pm\mp}\left(k^2\right)=\int \frac{d^2 p}{(2\pi)^2}\frac{4p^{\mp}(k-p)^{\pm}}{(p^2-m^2+i\varepsilon)((k-p)^2-m^2+i\varepsilon)}\,, \\
        & I_2^{\pm\pm}\left(k^2\right)=\int \frac{d^2 p}{(2\pi)^2}\frac{4p^{\pm}(k-p)^{\pm}}{(p^2-m^2+i\varepsilon)((k-p)^2-m^2+i\varepsilon)}\,. \notag
\end{align}
The difference now is that the integrals \eqref{2nd_group_loop_integrals} contain the logarithmic divergence and therefore have to be regularized ($\mu$ is the regularization parameter), so the calculation result is (as in the previous case details are written in the appendix \ref{app_loop_integrals})
\begin{align}
    & I_2^{+-}(s)=-\frac{i\theta + \pi}{4\pi}\left(\coth\frac{\theta}{2}-\frac{1}{\sinh\theta}\right)+\frac{i}{4\pi}\left(\frac{1}{\epsilon}-\gamma_E+\log(4\pi)-\log\frac{m^2}{\mu^2} \right)+ \mathcal{O}(\epsilon)\,, \\
    & I_2^{+-}(t)=\begin{cases}
    -\frac{i\theta}{4\pi}\left(\tanh\frac{\theta}{2}+\frac{1}{\sinh\theta}\right)+\frac{i}{4\pi}\left(\frac{1}{\epsilon}-\gamma_E+\log(4\pi)-\log\frac{m^2}{\mu^2} \right)+\mathcal{O}(\epsilon)\,, \, k_{3,4}=-k_{1,2}\,, \\
    \frac{i}{4\pi}\left(\frac{1}{\epsilon}-1-\gamma_E+\log(4\pi)-\log\frac{m^2}{\mu^2} \right)+\mathcal{O}(\epsilon)\,, \, k_4=-k_1\,, \, k_3 =-k_2\,,
    \end{cases} \notag \\    
    & I_2^{+-}(u) = 
    \begin{cases}
     \frac{i}{4\pi}\left(\frac{1}{\epsilon}-1-\gamma_E+\log(4\pi)-\log\frac{m^2}{\mu^2} \right)+\mathcal{O}(\epsilon)\,, \, k_4=-k_1\,, \, k_3=-k_2\,, \\
     -\frac{i\theta}{4\pi}\left(\tanh\frac{\theta}{2}+\frac{1}{\sinh\theta}\right)+\frac{i}{4\pi}\left(\frac{1}{\epsilon}-\gamma_E+\log(4\pi)-\log\frac{m^2}{\mu^2} \right)+\mathcal{O}(\epsilon)\,, \, k_{3,4}=-k_{1,2}\,,
    \end{cases} \notag
    \\
    & I_2^{\pm\pm}(s) 
    =-(k_1^{\pm}+k_2^{\pm})^2\left(\frac{i}{4\pi m^2\cosh^2\frac{\theta}{2}}+\frac{\pi+i\theta}{8\pi m^2\cosh^4\frac{\theta}{2}}\coth\frac{\theta}{2}
    \right)\,, \notag
    \\
    & I_2^{\pm\pm}(t)=
    \begin{cases}
    (k_1^{\pm}+k_4^{\pm})^2\left(\frac{ i}{4\pi m^2\sinh^2\frac{\theta}{2}}-\frac{ i\theta }{8\pi m^2\sinh^4\frac{\theta}{2}}\tanh\frac{\theta}{2}\right)\,,\,k_3 = -k_1,\,k_4=-k_2\,,
    \\
    0,\,k_3 = -k_2,\,k_4=-k_1\,. \notag
    \end{cases}
\end{align}

The third group are the diagrams consisting of all types of propagators, dashed, continuous and half-dashed half-continuous. The corresponding loop integral has the following form
\begin{equation}\label{3rd_group_loop_integrals}
        I_3^{\pm}(k) = \int \frac{d^2p}{(2\pi)^2}\frac{2imp^{\pm}}{(p^2-m^2+i\varepsilon)((k-p)^2-m^2+i\varepsilon)}\,.
\end{equation}
The integrals \eqref{3rd_group_loop_integrals} are convergent and for the cases we are interested in they give the result (see the details in the~\ref{app_loop_integrals})
\begin{align}
    & I_3^{\pm}(s) = -\frac{(k_1+k_2)^{\pm}}{4\pi m \sinh\theta}(i\pi - \theta)\,, 
    \\
    & I_3^{\pm}(t) =
    \begin{cases}
      -\frac{(k_1-k_2)^{\pm}\theta}{4\pi m \sinh\theta}\,, \quad k_3=-k_1\,, \quad k_4 = -k_2\,,
      \\
      0\,, \quad k_3=-k_2\,, \quad k_4=-k_1\,.
    \end{cases}\,.\notag
\end{align}
We notice that the u-channel integrals do not exist for the third group of diagrams.

The total one-loop contribution to the four-point correlation function is too cumbersome to be written here, but let us comment the disappearance of the divergence from the final result. It can be noticed that the divergence comes from the $I_2^{+-}$ group of diagrams. Indeed, the divergent contribution enters to the 4-point function (see Figure~\ref{1_loop_4_point_2nd_type}) with the coefficient
\begin{align}
    \frac{1}{4\pi\epsilon}\,4(im)^2 \left((k_3^-k_4^+ + k_1^-k_2^+) + (k_1^-k_4^+ + k_3^+k_4^-) - (k_1^-k_2^+ + k_1^-k_4^+ + k_3^-k_2^+ + k_3^-k_4^+)\right) = 0\,,
\end{align}
which proves the statement that the one-loop correlation function $\langle \tilde{\gamma}_1 \tilde{\bar{\gamma}}_2 \tilde{\gamma}_3 \tilde{\bar{\gamma}}_4 \rangle$ is convergent (of course, we take into account the counterterms proportional to $\delta_m^{(1)}$ \eqref{deltam1_counterterm}).

In the same way we can answer the question about the one loop renormalization of the quartic vertex in \eqref{Minkowski_betagamma_Larangian_renormalized}. Similarly, one can show that divergence of four-point function $\langle\beta_{+1}\beta_{-2}\gamma_3\bar\gamma_4\rangle$ also vanishes. Logarithmic divergence could appear from the diagrams of the $I_2^{+-}$ groups, which have the form as divergent diagrams for  $\langle\gamma_1\bar\gamma_2\gamma_3\bar\gamma_4\rangle$ up to replacement of  input propagators $(\gamma_1 \rightarrow \beta_{+1},\,\bar\gamma_2 \rightarrow \beta_{-2})$,
and enters to the correlation function with the coefficient
\begin{align}
    &\frac{1}{4\pi\epsilon}(16k_1^+k_2^-k_3^-k_4^+ + (im)^4 + 4(im)^2(k_2^-k_4^+ + k_1^+k_3^-) -
    \\
    &- (im)^4-4(im)^2(k_2^-k_4^+ + k_1^+k_3^-) 
    - 16k_1^+k_2^-k_3^-k_4^+)=0\,, \notag
\end{align}
where $k_1$, $k_2$ are the momenta of $\beta_+$, $\beta_-$ and $k_3$, $k_4$ are the momenta of $\gamma$, $\bar\gamma$ respectively. Thus, we do not need to introduce any $g^4$ order counterterms to the coupling constant $g^2$ at the one-loop level.
 
Applying the LSZ reduction formula and rewriting in terms of the rapidities we obtain the one-loop contribution to the $2 \rightarrow 2$ amplitudes for $\gamma$ and $\bar{\gamma}$ in the form
\begin{equation}\label{1_loop_gammagammabar_amplitude}
    \mathcal{M}_{\gamma\bar{\gamma}}^{\gamma\bar{\gamma}(1-\text{loop})}(\theta)=\begin{cases}
        -\frac{g^4m^4}{\pi}\sinh^2 \frac{\theta}{2} \left(i+\pi\tanh\frac{\theta}{2}\right)\,, \quad k_3=-k_2\,, \quad k_4=-k_1\,, \\
        0\,, \quad k_3=-k_1\,, \quad k_4=-k_2\,.
       \end{cases} 
\end{equation}
Thus, we observe the reflectionless scattering and, therefore, the Yang-Baxter equation is satisfied at one loop as well. Now we can extract the contribution to the element of $S$-matrix by applying the formula \eqref{2_particle_S_matrix} to \eqref{1_loop_gammagammabar_amplitude}. Eventually, the transition amplitude from \eqref{1_loop_gammagammabar_amplitude} gives us exactly the same one-loop contribution as in \eqref{eq:Seaxt2}. Therefore, by combining \eqref{tree_level_gammagammabar_amplitude} with \eqref{1_loop_gammagammabar_amplitude}, we can write the $S$-matrix element of $\gamma\bar{\gamma} \rightarrow \gamma\bar{\gamma}$ scattering in our theory
\begin{align}
    & S_{\gamma\bar{\gamma}}^{\gamma\bar{\gamma}}(\theta) = 1 +\frac{ig^2}{2}( \cosech\theta-\coth\theta) +\frac{ig^4}{8\pi}(\cosech\theta - \coth\theta)+ \\
    & +\frac{g^4}{2}\coth\theta\,\cosech\theta -\frac{g^4}{2}\left( \frac{1}{2}+\cosech^2\theta\right) + \mathcal{O}(g^6)\,. \notag
\end{align}
which exactly coincides with the one \eqref{eq:Seaxt2} from \cite{Dorey:1994mg}, but differs at the one loop order with \eqref{eq:S-matrix} from \cite{deVega:1981ka}, which can be attributed to the difference in the details of renormalization schemes, as it was noticed in \cite{Hoare:2010fb}.

In addition, we recall that in the $\beta\gamma$ action \eqref{betagamma_action_gWZW} originating from the gWZW model the constant $k$ is related to the coupling as
\begin{equation}\label{betagamma_gWZW_coupling}
    g^2=\frac{4\pi}{k}\,.
\end{equation}
Taking into account \eqref{coupling_quantization} we arrive to the relation
\begin{equation}\label{kk_relation_expansion}
\frac{1}{\mathrm{k}}=\frac{1}{k}+\frac{1}{k^2}+\mathcal{O}\left(\frac{1}{k^3}\right)\,.
\end{equation}
Given \eqref{kk_relation_expansion} it is natural to assume that the formula \eqref{coupling_constant_renormalization} for the relation between $g^2$ and $g_R^2$ is equivalent to the relation
\begin{equation}\label{k_shift}
    k=\mathrm{k}-1\,.
\end{equation}
Checking the relation \eqref{k_shift} at the higher orders in $k$ will require calculating the S-matrix at the two-loop order and higher, which is the subject of future work.

Let us summarize our results and discuss possible open problems in the conluding section of our work.

\section{Conclusions}\label{Conclusions}

In this article we conducted analysis of the version of bosonic Thirring model in Minkowski spacetime. We started from remembering the relation of the Euclidean version of this model to the complex sine-Gordon model via the use of path integral. Then we offered a way to analytically continue the bosonic version of Thirring model to the Minkowski signature, putting forward the hypothesis that such a definition of this theory allows to reproduce the $S$-matrix of the Minkowski version of complex sine-Gordon model from \cite{Dorey:1994mg}. Also we managed to transform the gauged Wess-Zumino-Witten model with integrable potential into this bosonic Thirring model in Minkowski signature ($\beta\gamma$-system) via field redefinitions.

To achieve the goal of reproducing the S-matrix we conducted some perturbative analysis of the $\beta\gamma$-system \eqref{Minkowski_betagamma_Lagrangian}. First of all we calculated the two-point correlation function of the theory at one loop order in the Subsection~\ref{1_loop_2_point_functions} and showed that it is enough to have only one loop mass renormalization terms to absorb the divergences appearing. Then we also found the two-loop two-point function, showing that there are three types of diagrams contributing. It appears that the one loop counterterms from the mass renormalization are sufficient to compensate all the divergences except for the ones coming from the so-called sunset diagrams. The divergences from the sunset diagrams can be absorbed into the nex-to-leading order counterterm to the mass terms in the Lagrangian.

In the next part we concentrated on calculating the amplitudes of the scattering of $\gamma$ and $\bar{\gamma}$ in order to compare the result with the $S$-matrix of the complex sine-Gordon theory. By computing the four-point function $\langle\tilde{\gamma}_1 \tilde{\bar{\gamma}}_2 \tilde{\gamma}_3 \tilde{\bar{\gamma}}_4 \rangle$ and applying the LSZ reduction formula we managed to find the corresponding $\gamma\bar{\gamma} \rightarrow \gamma\bar{\gamma}$ $S$-matrix element up to one loop order. And indeed it coincides with the same $S$-matrix element from \cite{Dorey:1994mg} at this perturbation theory order, which represents the argument in favour of our initial hypothesis that the theory \eqref{Minkowski_betagamma_Lagrangian} correctly reproduces the $S$-matrix of the complex sine-Gordon theory. The one-loop result from \cite{deVega:1981ka} differs from ours and the one from \cite{Dorey:1994mg}, however, this could be attributed to the details of the renormalization scheme used, as it was noticed in \cite{Hoare:2010fb}. As a byproduct we checked the the one-loop corrections to the quartic vertex $\beta_+ \beta_- \bar{\gamma} \gamma$ do not have any divergences, therefore confirming that there is no need for the counterterms to the coupling constant at the order $g^4$.

An obvious direction for future study is the analysis of the theory \eqref{Minkowski_betagamma_Lagrangian} at higher loops. It would be important to proceed first of all with the study of renormalizability properties. The next step would be to find the two-loop contribution to the 4-point correlation function aiming to extract from it the two-loop amplitude, which can compared with the all-loop result of \cite{Dorey:1994mg}. This can shed some light on which counterterms besides the mass ones are needed at the two-loop order.

Another interesting problem is to consider several copies of the $\beta\gamma$-system with the interaction term proportional to $r_{ijkl}\bar{\beta}^{(i)}\beta^{(j)}\bar{\gamma}^{(k)}\gamma^{(l)}$, where the indices $(i)$ denote the isotopic index of the field, as it was done in \cite{Bykov:2019vkf,Bykov:2020nal}. It was shown there that a theory with such an interaction is classically integrable provided that $r_{ijkl}$ satisfies classical Yang-Baxter equation. One can pose a question whether this condition is sufficient for the quantum integrability of the Minkowskian version of the model with such an interaction. The key test for the quantum integrability of such a coupled $\beta\gamma$-system would be to check the absence of particle production \cite{Dorey:1996gd}.

To sum up, the result obtained in this article together with the mentioned directions of future study could improve the understanding of the CSG model itself and help to progress with the dual description of deformed $OSp(N|2m)$, as well as other integrable sigma models, as it represents an important ingredient in the calculation of the $S$-matrix of the dual Toda-like theory.

\section*{Acknowledgements}

M.A. and A.K. are especially grateful to Ben Hoare for illuminating ideas and participating in the work at various stages of the project. M.A. would also thank Alexey Litvinov, Arkady Tseytlin, Mikhail Vasiliev and Dmitri Bykov for fruitful and stimulating discussions and Anna Popova for the inspiration and linguistic assistance during the creation of this text.

\appendix

\section{Path integral in the Euclidean $\beta\gamma$-system}\label{app_Path_integral}

In order to demonstrate the correspondence between the $\beta\gamma$-system and the CSG defined in Euclidean space we are going calculate the path integral over $\beta$ and $\bar\beta$ fields
\begin{align}\label{eq:PathInt}
        & \int \mathcal{D}\beta \mathcal{D}\bar\beta \exp\left[-\int dxdy\; \mathcal{L}^{(E)}_{\beta\gamma}\right] = \exp\left[\int dxdy\; m\gamma\bar\gamma\right]\times
        \\
        & \times \int \mathcal{D}\beta \mathcal{D}\bar\beta \exp\left[-\int dxdy \left( \beta\bar{\partial}\gamma+\bar{\beta}\partial\bar{\gamma}-m \beta\bar{\beta}+g^2 \beta\bar{\beta}\gamma\bar{\gamma}\right)\right]\,.\notag
\end{align}

Let us re-define the fields $\beta$ and $\bar\beta$ by making the following shifts
\begin{equation}\label{beta_field_redefinition}
    \beta \rightarrow \beta' = \beta + \frac{\partial\bar \gamma}{g^2\gamma\bar\gamma - m}\,, \quad \bar\beta \rightarrow \bar\beta' = \bar\beta + \frac{\bar\partial \gamma}{g^2\gamma\bar\gamma - m}\,. 
\end{equation}
Substituting the redefined fields \eqref{beta_field_redefinition} into the path integral \eqref{eq:PathInt} we obtain the integral of the Gaussian form 
\begin{align}
    & \int \mathcal{D}\beta' \mathcal{D}\bar\beta' \exp\left[-\int dxdy \; \left(\beta'(g^2\gamma\bar\gamma -m)\bar\beta' - m\gamma\bar\gamma +\frac{\bar\partial \gamma\partial\bar \gamma}{m - g^2\gamma\bar\gamma}\right)\right]=
    \\
    & =\frac{\pi}{\det(m-g^2\bar\gamma\gamma )} \exp\left[-\int dxdy \left( \frac{\bar\partial \gamma\partial\bar \gamma}{m - g^2\gamma\bar\gamma} - m\gamma\bar\gamma\right)\right]\,. \notag
\end{align}
If we pass to the massless fields $\gamma'$ and $\bar{\gamma}'$
\begin{equation}
    \gamma \rightarrow \gamma' = \frac{\gamma}{\sqrt{m}}\,, \quad \bar{\gamma} \rightarrow \bar{\gamma}' = \frac{\bar{\gamma}}{\sqrt{m}}\,,
\end{equation}
then in the exponent we obtain the classical CSG Lagrangian
\begin{align}
    \frac{1}{\det(1-g^2\bar\gamma'\gamma' )} \exp\left[-\int dxdy \left( \frac{\bar\partial \gamma'\partial\bar \gamma'}{1-g^2\gamma'\bar\gamma'}-m^2\gamma'\bar\gamma'\right)\right]\,.
\end{align}
Thus, taking path integral \eqref{eq:PathInt} we get the classical CSG theory, defined in Euclidean space, modulo the contribution of the functional determinant.

Let us show that the imaginary part of the CSG Lagrangian defined in Minkowski space is a total derivative and can be ignored and, thus, Lagrangian can be considered as a real function. Consider the imaginary part of the term
\begin{align}\label{eq:derivative}
    \frac{\partial_+\bar\gamma\partial_-\gamma}{1-g^2\gamma\bar\gamma},
\end{align}
where the fields $\gamma$ and $\bar{\gamma}$ can be written in the form: $\gamma = re^{i\phi},\;\bar\gamma = re^{-i\phi}$. Then, for the imaginary part of \eqref{eq:derivative} we obtain
\begin{equation}\label{Im_to_polar}
\Im\left(\frac{\partial_+\bar\gamma\partial_-\gamma}{1-g^2\gamma\bar\gamma}\right) = \frac{r}{1-g^2r^2}\left( \partial_+r\;\partial_-\phi-\partial_+\phi\;\partial_-r)\right) =\partial_+f(r)\;\partial_-\phi- \partial_+\phi\;\partial_-f(r)\,,
\end{equation}
where we introduced the notation
\begin{equation}\label{f(r)_definition}
    f(r) = -\frac{1}{2g^2}\log(1-g^2r^2)\,.  
\end{equation}
Integraing by parts and flipping the derivatives under the integral, for the imaginary part of the term \eqref{Im_to_polar} we get
\begin{align}\label{integration_Im_by_parts}
\int \,dx\,dy \Im\left(\frac{\partial_+\bar\gamma\partial_-\gamma}{1-g^2\gamma\bar\gamma}\right) = \int \,dx\,dy  (-f(r)\partial_+\partial_-\phi+f(r)\partial_-\partial_+\phi) = 0 \,,    
\end{align}
where we took into account that the fields vanish at infinity. Thus, the imaginary part of CSG Lagrangian in Minkowski space is total derivative and we can consider it as a real Lagrangian.

\section{Loop integrals calculation}\label{app_loop_integrals}

The present Appendix is devoted to the details of the calculation of loop integrals appearing in the process of deriving the two- and four-point functions of the theory in question.

\subsection{$I_1$ calculation}\label{app_I1_integral}

Let us calculate loop integral $I_1(k^2)$. This integral is similar to the well-known loop integral for the theory $\phi^4$ and equals to following
\begin{equation}
I_1(k^2)=-\int\frac{d^2 p}{(2\pi)^2}\frac{m^2}{(p^2-m^2+i\varepsilon)((k-p)^2-m^2+i\varepsilon)} =
-\frac{m^2}{2\pi k^2}\sqrt{\frac{k^2}{4m^2-k^2}}\log\frac{1+i\sqrt{\frac{k^2}{4m^2-k^2}}}{1-i\sqrt{\frac{k^2}{4m^2-k^2}}}\,. 
\end{equation}

\subsection{$I_3$ calculation}\label{app_I2_integral}

Let us calculate the loop integral $I_3^{\pm}(k)$ corresponding to the loops formed by the dashed or continuous line and half-dashed  half-continuous line
\begin{equation}
     I_3^{\pm}(k) = \int \frac{d^2p}{(2\pi)^2}\frac{2imp^{\pm}}{(p^2-m^2+i\varepsilon)((k-p)^2-m^2+i\varepsilon)}\,.
\end{equation}
On the other hand, the integral can be rewritten as
\begin{equation}
     I_3^{\pm}(k) = \int \frac{d^2p}{(2\pi)^2}\frac{2im(k-p)^{\pm}}{(p^2-m^2+i\varepsilon)((k-p)^2-m^2+i\varepsilon)}\,.
\end{equation}
Comparing both expressions, we obtain the expression for the integral $I_3^{\pm}(k)$  through the integral $I_1(k^2)$
\begin{equation}
    I_3^{\pm}(k) = -\frac{2ik^{\pm}}{m}I_1(k^2) -I_3^{\pm}(k) \rightarrow I_3^{\pm}(k) = \frac{ik^{\pm}}{m}I_1(k^2)\,.
\end{equation}
Thus, we derive the following expression for the loop integral $I_3^{\pm}(k)$
\begin{equation}
    I_3^{\pm}(k)=\frac{im  k^{\pm}}{2\pi k^2}\sqrt{\frac{k^2}{4m^2-k^2}}\log\frac{1+i\sqrt{\frac{k^2}{4m^2-k^2}}}{1-i\sqrt{\frac{k^2}{4m^2-k^2}}}\,.
\end{equation}

\subsection{$I_2$ calculation}\label{app_I3_integral}

Let us calculate the loop integral $I_2^{\pm}(p)$ corresponding to the loops formed by two mixed lines
\begin{align}
& I^{+-}_2(k^2) = \int\frac{d^2p}{(2\pi)^2}\frac{4p^{\mp}(k-p)^{\pm}}{(p^2-m^2+i\varepsilon)((k-p)^2-m^2+i\varepsilon)} = \\
& =\int \frac{d^2p}{(2\pi)^2}\left(\frac{4p^{\mp}k^{\pm}}{(p^2-m^2+i\varepsilon)((k-p)^2-m^2+i\varepsilon)}-\frac{p^2}{(p^2-m^2+i\varepsilon)((k-p)^2-m^2+i\varepsilon)}\right) = \notag
\\
&=I_{2a}^{+-}(k^2) - I_{2b}^{+-}(k^2)\,, \notag
\end{align}
where take into account that $p^2 = 4p^+p^{-}$.

Consider the contributions of terms to the $I_2^{+-}(k^2)$ separately. Let us take the second integral (see the calculation of $I_3(k)$)
\begin{equation}
    I^{+-}_{2a} (k^2)= \frac{1}{4\pi}\sqrt{\frac{k^2}{4m^2-k^2}}\log\frac{1+i\sqrt{\frac{k^2}{4m^2-k^2}}}{1-i\sqrt{\frac{k^2}{4m^2-k^2}}}\,.
\end{equation}    
The second contribution has logarithmic divergence. We regularized this integral using dimensional regularization \footnote{In calculating the integrals we carried out the Wick rotation defined as $p^0 \xrightarrow[]{Euclid} ip_E^0\,,\,d^2p\xrightarrow[]{Euclid}id^2p_E\,,\, (p_E,x_E) = -2(p_E^+x_E^- + p_E^-x_E^+)\,,\; p_E^2 = -4p_E^+ p_E^- = -p^2\,.$}
\begin{align}
    &I_{2b}^{+-}(k^2)=\int\frac{d^2p}{(2\pi)^2}\frac{p^2}{(p^2-m^2+i\varepsilon)((k-p)^2-m^2+i\varepsilon)}
    \xrightarrow[]{Euclid} 
    \\
    &\xrightarrow[]{Euclid} -i\int \frac{\mu^{2\epsilon}d^{2-2\epsilon}p_E}{(2\pi)^{2-2\epsilon}}\,\frac{1}{(k-p)_E^2+m^2} - I_1(k_E^2)\xrightarrow[]{Mink}\frac{m^2}{2\pi k^2}\sqrt{\frac{k^2}{4m^2-k^2}}\log\frac{1+i\sqrt{\frac{k^2}{4m^2-k^2}}}{1-i\sqrt{\frac{k^2}{4m^2-k^2}}} -\notag
    \\
    &-\frac{i}{4\pi}\left(\frac{1}{\epsilon}-\gamma_E+\log(4\pi)-\log\frac{m^2}{\mu^2} + \mathcal{O}(\epsilon)\right)\,. \notag
\end{align}

Therefore, the one-loop integral $I_2^{+-}(k^2)$ takes the form
\begin{align}
    &I_2^{+-}(k^2) =\frac{1}{4\pi}
   \left(1-\frac{2m^2}{k^2}\right)\sqrt{\frac{k^2}{4m^2-k^2}}\log\frac{1+i\sqrt{\frac{k^2}{4m^2-k^2}}}{1-i\sqrt{\frac{k^2}{4m^2-k^2}}}- 
   \\
   &+\frac{i}{4\pi}\left(\frac{1}{\epsilon}-\gamma_E+\log(4\pi)-\log\frac{m^2}{\mu^2} + \mathcal{O}(\epsilon)\right)\,. \notag
\end{align}

Let us calculate the integral $I_2^{\pm\pm}(k)$
\begin{align}
    &I_2^{\pm\pm}(k) = \int \frac{d^2p}{(2\pi)^2} \frac{4p^{\pm}(k-p)^{\pm}}{(p^2-m^2+i\varepsilon)((k-p)^2-m^2+i\varepsilon)}\xrightarrow[]{Euclid}\int \frac{d^2p_E}{(2\pi)^2} \frac{4ip_E^{\pm}(k-p)_E^{\pm}}{(p_E^2+m^2)((k-p)_E^2+m^2)}= 
    \\
    &= \int\limits_0^1 dx\,\int \frac{d^2p_E}{(2\pi)^2} \frac{4ip_E^{\pm}(k-p)_E^{\pm}}{(p_E^2-2(p,k)_Ex+k_E^2x + m^2)^2} =\int\limits_0^1 dx\,\int \frac{d^2p_E}{(2\pi)^2} \frac{4i(p_E^{\pm}+k_E^{\pm}x)(k_E^{\pm}(1-x)-p_E^{\pm})}{(p_E^2+k_E^2x(1-x) + m^2)^2} = \notag
    \\
    &=\int\limits_0^1 dx\,\int \frac{d^2p_E}{(2\pi)^2} \frac{4i (k_E^{\pm})^2x(1-x)}{(p_E^2+k_E^2x(1-x) + m^2)^2} =\frac{i(k_E^{\pm})^2}{\pi k_E^2} + \frac{2m^2(k_E^{\pm})^2}{\pi k_E^4}\sqrt{\frac{k_E^2}{-4m^2-k_E^2}}\log\frac{1+i\sqrt{\frac{k_E^2}{-4m^2-k_E^2}}}{1-i\sqrt{\frac{k_E^2}{-4m^2-k_E^2}}}\,.\notag
\end{align}

Thus, the integral $I_2^{\pm\pm}(k)$ in Minkowski signature has the form
\begin{equation}
    I_2^{\pm\pm}(k)=-\frac{ i (k^{\pm})^2}{\pi k^2} + \frac{2 m^2(k^{\pm})^2}{\pi k^4}\sqrt{\frac{k^2}{4m^2-k^2}}\log\frac{1+i\sqrt{\frac{k^2}{4m^2-k^2}}}{1-i\sqrt{\frac{k^2}{4m^2-k^2}}}\,.
\end{equation}

\subsection{$J_{1,2,3}$ integral calculation}\label{J_integrals_indetities}

The integral $J_0(k^2)$ is similar to the sunset integral for the $\phi^4$ theory \cite{Kleinert:2001ax}
\begin{align}
    &J_0(k^2) = -\int\frac{d^2p_1}{(2\pi)^2}\frac{d^2p_1}{(2\pi)^2}\,\frac{im^3}{(p_1^2-m^2+i\varepsilon)(p_2^2-m^2+i\varepsilon)((k-p_1-p_2)^2-m^2+i\varepsilon)} = 
    \\
    &=-im^3(3m^2A(k^2) + B(k^2))\,,\notag
\end{align}
where
\begin{align}
    &A(k^2) = \frac{\Gamma(2)}{(4\pi)^2}\int\limits_0^1 dx\, \frac{1}{x(1-x)}\int\limits_0^1 dy\,\frac{y}{\left(-k^2y(1-y)+m^2\left(y+\frac{1-y}{x(1-x)}\right)\right)^2}\,,
    \\
    &B(k^2) = -k^2\frac{\Gamma(1)}{(4\pi)^2}\int\limits_0^1dx\,\frac{1}{x(1-x)}\int\limits_0^1 dy\,y(1-y)\frac{1}{\left(-k^2 y(1-y)+m^2\left(1-y+\frac{y}{x(1-x)}\right)\right)^2}\,.\notag
\end{align}

The integral $J_1(k)$ can be expressed in terms of the integral $J_0(k^2)$
\begin{equation}
     J_1^{\pm}(k)=-\frac{2i}{3}\frac{k^{\pm}}{m}J_0(k^2)\,.
\end{equation}
Similar procedure for the integral $J_2^{+-}(k^2)$ yields
\begin{equation}
    J_2^{+-}(k^2)=-\frac{k^2-3m^2}{6m^2}J_0(k^2)+\frac{i}{2m}\left(\int\frac{d^2 p}{(2\pi)^2}\frac{im}{p^2-m^2}\right)^2\,.
\end{equation}

The third sunset integral can be calculated using the Feynman parameter technique
\begin{align}
    J_3^{\pm\pm\mp}(k) = \int\frac{d^2p_1}{(2\pi)^2}\frac{d^2p_1}{(2\pi)^2}\,\frac{8p_1^{\pm}p_2^{\pm}(k-p_1-p_2)^{\mp}}{(p_1^2-m^2+i\varepsilon)(p_2^2-m^2+i\varepsilon)((k-p_1-p_2)^2-m^2+i\varepsilon)}\,.
\end{align}

Consider individual contributions to the integral $J_3^{\pm\pm\mp}(k)$ integral. Let us calculate the following term
\begin{align}
    &J_a(k) = \int\frac{d^2p_{1E}}{(2\pi)^2}\frac{d^2p_{2E}}{(2\pi)^2}\frac{8k^{\mp}p_1^{\pm}p_2^{\pm}}{(p_1^2 - m^2+i\varepsilon)(p_2^2 - m^2+i\varepsilon)((k-p_1-p_2)^2-m^2+i\varepsilon)} \xrightarrow[]{Euclid}
    \\
    &\xrightarrow[]{Euclid}\int\frac{d^2p_{1E}}{(2\pi)^2}\frac{d^2p_{2E}}{(2\pi)^2}\frac{8k_E^{\mp}p_{1E}^{\pm}p_{2E}^{\pm}}{(p_{1E}^2 + m^2)(p_{2E}^2 + m^2)((k-p_{1}-p_{2})_E^2+m^2)} = \notag
    \\    &=8k_E^{\mp}\int\frac{d^2p_{2E}}{(2\pi)^2}\frac{p_{2E}^{\pm}}{p^{2}_{2E}+m^2}\int\limits_0^1 dx\int\frac{d^2p_{1E}}{(2\pi)^2}\frac{p_{1E}^{\pm}}{(p_{1E}^2-2(p_{1},q)_Ex + q^2x + m^2)^2} = \notag
    \\
    &= 8k_E^{\mp}\int\limits_0^1 dx\,\frac{1}{4\pi(1-x)}\int\limits_0^1 dy\int\frac{d^2p_{2E}}{(2\pi)^2}\frac{p_{2E}^{\pm}(k-p_2)_E^{\pm}}{\left(k_{2E}^2 - 2(p_2,k)_Ey +k_E^2y + M^2(x,y)\right)^2} \,,\notag
\end{align}
where we defined the function depending only on Feynman parameters
\begin{equation}
    M^2(x,y) = \frac{ym^2}{x(1-x)} + m^2 - m^2y\,.
\end{equation}
The integral over the loop momentum $p_{2E}$ is equal to the expression
\begin{align}
    &\int\frac{d^2p_{2E}}{(2\pi)^2}\frac{p_{2E}^{\pm}k_E^{\pm}-(p_{2E}^{\pm})^2}{\left(p_{2E}^2 - 2(p_2,k)_E y +k_E^2y + M^2(x,y)\right)^2} = \int\frac{d^2p_{2E}}{(2\pi)^2}\frac{(p_{2E}^{\pm} + k_E^{\pm}y)(k_E^{\pm}(1-y)-p_{2E}^{\pm})}{\left(p_{2E}^2 + k^2_Ey(1-y) +  M^2(x,y)\right)^2}=
    \\
    &=\int\frac{d^2p_{2E}}{(2\pi)^2}\frac{(k_E^+)^2\,y(1-y)}{\left(p_{2E}^2 + k^2_Ey(1-y) +  M^2(x,y)\right)^2} = \frac{(k_E^+)^2y(1-y)}{4\pi}\frac{1}{k_E^2y(1-y)+M^2(x,y)}\,.\notag
\end{align}

Therefore,  $J_a(k)$ in Minkowski signature has the following form
\begin{align}
    &\frac{8k^-(k^+)^2}{16\pi^2}\int\limits_0^1dx \int\limits_0^1dy \,\frac{y(1-y)}{1-x}\frac{1}{-k^2y(1-y)+M^2(x,y)}
\end{align}

Consider the contribution to $J_3^{\pm\pm\mp}(k)$
\begin{align}
    &J_b(k)=-\int\frac{d^2p_{1}}{(2\pi)^2}\frac{d^2p_{2}}{(2\pi)^2}\frac{2p_{1}^2p_{2}^{\pm}}{(p_{1}^2 - m^2+i\varepsilon)(p_{2}^2 - m^2+i\varepsilon)((k-p_1-p_2+i\varepsilon)^2-m^2)} =\\
    & = -\underbrace{\int\frac{d^2p_{1}}{(2\pi)^2}\frac{d^2p_{2}}{(2\pi)^2}\frac{2p_{2}^{\pm}}{(p_{2}^2 - m^2)((k-p_1-p_2)^2-m^2)}}_{0} + J_1^+(k)\,.\notag
\end{align}
Taking into account the symmetry of $J_3^{\pm\pm\mp}(k)$ with respect to the replacement $p_1 \leftrightarrow p_2$ the given contribution $J_b(k)$ enters into $J_3^{\pm\pm\mp}(k)$ twice.

Thus we obtain the following expression for integral $J_3^{\pm\pm\mp}(k)$ in Minkowski signature
\begin{align}
    J_3^{\pm\pm\mp}(k) = \frac{k^{\pm}(k)^2}{8\pi^2}\int\limits_0^1dx\,\frac{1}{1-x} \int\limits_0^1dy \,\frac{y(1-y)}{-k^2y(1-y)+M^2(x,y)} + 2J_1^{\pm}(k) \,.
\end{align}

\bibliographystyle{MyStyle}
\bibliography{MyBib}

\end{document}